\documentclass[twocolumn,aps,rmp]{revtex4}
\usepackage{amsfonts,amssymb,amsmath,bm}
\usepackage[dvips]{graphicx}
\usepackage{times}
\usepackage{verbatim}
\usepackage{theorem}
\usepackage{makeidx}
\usepackage{amsmath}
\usepackage{epic}
\usepackage{wasysym}
\usepackage{amscd}
\usepackage{bbm}
\usepackage{xy}
\usepackage{amssymb}
\usepackage{epsfig}

\begin{document}

\title{Principles of the motion of fluids} 
\thanks{This is an English adaptation by Walter Pauls 
of Euler's memoir `Principia motus fluidorum' (Euler, 1756--1757). Updated
versions of the translation may become available at 
\url{www.oca.eu/etc7/EE250/texts/euler1761eng.pdf}.
  For a detailed presentation
of Euler's fluid dynamics papers, cf.  Truesdell, 1954, which has
also been helpful for this translation. Euler's
work is discussed in the perspective of eighteenth century
fluid dynamics research by Darrigol and
Frisch, 2008. The help of O.~Darrigol, U.~Frisch, G.~Grimberg and G.~Mikhailov
is also acknowledged.\\
Explanatory footnotes and references have been supplied where necessary; Euler's memoir had
 neither
  footnotes nor a list of references.}
\author{Leonhard Euler}

\begin{abstract}

Here are treated the elements of the theory of the
motion of fluids in general, the whole matter being reduced to
this: given a mass of fluid, either free or confined in
vessels, upon which an arbitrary motion is impressed,
and which in  turn is acted upon by arbitrary forces, to determine
the motion carrying forward each particle, and at the same time to
ascertain the pressure exerted by each part, acting on it as
well as on the sides of the vessel. At first in
this  memoir, before undertaking the
investigation of  these effects of the forces,  the Most Famous
Author\footnote{Summaries, which at that time were not placed at the
beginning of the corresponding paper, were published under the responsibility 
of the Academy; the presence  of the  words ``Most Famous Author'', rather
common at the time, 
cannot be taken as  evidence that Euler usually referred to himself in 
this way.} carefully
evaluates all the possible motions which can actually take place in the
fluid. Indeed, even if the individual particles of the fluid are free 
from each other,  motions in which the particles interpenetrate 
are nevertheless excluded, since we are dealing with fluids that do not permit 
any compression into a narrower volume.
Thus it is clear that an arbitrary small
portion of fluid cannot receive a motion other than the one which
constantly conserves the same volume; even though meanwhile the shape
is changed in any way. It would hold indeed, as long as  no
elementary portion would be compressed  at any time  into a smaller
volume; furthermore\footnote{In the original, we find ``verum quoniam''; the
litteral translation ``since indeed'' does not seem  logically consistent. }
if the portion expanded into a larger 
volume, the continuity of the particles was violated, these were dispersed
and no longer clinged together, such a motion would no longer
pertain to the science of the motion of fluids; but individual
droplets would separately perform their motion.  Therefore, this case
being excluded, the motion of the fluids must be restricted by this
rule that each small portion must retain for ever the same volume;
and this principle  restricts the general expressions of motion for
elements of the fluid. Plainly, considering an
arbitrary small portion of the fluid, its individual points have to be
carried by such a motion that, when at a moment of time they arrive at
the next location, till then they occupy a volume equal to the
previous one; thus if, as usual, the motion of a point is decomposed
parallel to fixed orthogonal directions, it is necessary that a
certain established relation hold  between these three velocities,
which the Author has determined in the first part.\\

In the second part the author proceeds to the determination of the
motion of a fluid produced by arbitrary forces, in which matter the
whole investigation reduces to this that the pressure with which the
parts of the fluid at each point act upon one another shall be
ascertained;  which pressure is denoted  most conveniently, as
customary for water, by a certain height; this is to be understood
thus, that the each  element of the fluid sustains a pressure the same
as if were pressed by a heavy column of the same fluid, whose
height is equal to that amount. Thus, in such way in each point of
the fluid the height referring to the state of the pressure will be
given;  since it is not equal to the one in the neighbourhood,
it will perturb the motion of the elements. But this pressure depends
as well on the forces acting on each element of the fluid,
as on those, acting in the whole mass; thus,  by the given forces,
the pressure in each point and thereupon the
acceleration of each element --  or its  retardation --  can be
assigned for the motion, all which determinations are expressed by the
author through differential formulae. But, in fact, 
the full development of these formulas mostly involves  the greatest
difficulties. But nevertheless this whole theory has been reduced to
pure analysis, and what remains to be completed in it depends solely
upon subsequent progress in Analysis. Thus it is far from true that
purely analytic researches are of no use in applied mathematics;
rather, important additions in pure analysis are now required.

\end{abstract}

\date{\today}

\maketitle

\section{First part}
\label{s:parsprior}

 {\bf 1}.~~~Since liquid substances differ from solid ones by the fact
that their particles are mutually independent of each other, they
can also receive most diverse motions; the motion performed by an
arbitrary particle of the fluid is not determined by the motion of the
remaining particles to the point that it cannot move in any other
way. The matter is very different in solid bodies,
which, if they were inflexible, would not undergo any change in
their shape; in whatsoever way they be moved, each of their
particles would constantly keep the same location and distance with
respect to other particles; it thus follows that, the motion of two or, if
necessary, three of all the particles being known, the motion of any
other particle can be defined; furthermore the motion of two
or three particles of such a body cannot be chosen at will, but must
be constrained in such a way that these particles preserve
constantly their positions with respect to each other.\footnote{Here
Euler refers to the motion of rigid solid bodies treated previously in
Euler, 1750.}

 {\bf 2}.~~~But if, moreover, solid bodies are flexible, the motion of
each particle is less constrained: because of the bending,
the distance as well as the relative position of each particle
can be subject to changes. However, the manner
itself of the bending constitutes a certain law which various
particles of such a body have to obey in their motion: certainly what
has to be taken care of is that the parts that experience in their
neighborhood such a strong bending with respect to each other are
neither torn apart from the inside nor penetrate into each other.
Indeed, as we shall see, impenetrability is demanded for all bodies.

{\bf 3}.~~~In fluid bodies, whose particles are
united among themselves by no bond, the motion of each
particle is much less restricted: the motion of the remaining
particles is not determined from the motion of any number of 
particles. Even knowing  the motion of one hundred particles, 
the future motion permitted to the remaining 
particles still can vary in infinitely many
ways. From which it is seen that the motion of these
fluid particles plainly does not depend on the motion of the
remaining ones, unless it be enclosed by these so that it is
constrained to follow them.
 
{\bf 4}.~~~However, it cannot happen that the motion of all
particles of the fluid suffers no restrictions at all. Furthermore,
one cannot at will invent a motion that is
conceived to occur for each particle. Since, indeed, the particles
are impenetrable, it is immediately clear that a motion cannot be
maintained in which some particles go through other particles and,
accordingly, penetrate each other: also, because of this reason such
motion certainly cannot be conceived to occur in the fluid.  Therefore,
infinitely many motions must be excluded; after their determination 
the remaining ones are grouped together.
It is seen worthwhile to define them more accurately regarding the
property which distinguishes them from the previous ones.

{\bf 5}.~~~But before the motion by which the fluid is agitated at any
place can be defined, it is necessary to see how every motion, which
can definitely be maintained in this fluid, be recognized: these
motions, here, I will call possible, which I will distinguish from
impossible motions which certainly cannot take place. We must then
find what characteristic is appropriate to possible motions,
separating them from impossible ones. When this is done, we shall have
to determine which one of all possible motions in a certain case ought
actually to occur. Plainly we must then turn to the forces which act
upon the water, so that the motion appropriate to them may be
determined from the principles of mechanics.

{\bf 6}.~~~Thus, I decided to inquire into the character of the
possible motions, such that no violation of the
impenetrability can occur in the fluid. I shall assume the fluid to be
such as never to permit itself to be forced into a lesser space, nor should
its continuity be interrupted. Once the theory of fluids has been
adjusted to fluids of this nature, it will not be difficult to extend
it also to those fluids whose density is variable and which do not
necessarily require continuity.\footnote{See the English translation of
"General laws of the motion of fluids" in these Proceedings.}

{\bf 7}.~~~If, thus, we consider an arbitrary portion in such a
fluid, the motion, by which each of its particles is carried has to
be set up  so that at each time they occupy an equal
volume. When this occurs in separate portions, any expansion
into a larger volume, or compression into a smaller volume is
prohibited. And, if we turn 
attention to this only property,  we can have only such motion that
the  fluid is  not permitted to expand or compress. Furthermore, what
is said  here about arbitrary portions of the fluid, has to be
understood for each of its elements; so that the volume of its
elements must  constantly preserve its value.

{\bf 8}.~~~Thus, assuming that this condition holds, let an arbitrary
motion be considered to occur at each point of the fluid; moreover,
given any element of the fluid, consider the brief translations of each of
its boundaries. In this manner the volume, in
which the element is contained after a very  short time, becomes
known. From there on, this volume is posed to be equal to the one
occupied previously, and this equation will prescribe the calculation of the
motion, in so far as it will be possible. Since all elements occupy
the same volumes during all periods of time, no compression of
the fluid, nor expansion can occur; and the motion is arranged in such
a way that this becomes possible.

{\bf 9}.~~~Since we consider not only the velocity\footnote{Meaning here the
absolute value of the velocity.} of the motion 
occuring at every points of the fluid but also its direction, both
aspects are most conveniently handled,
if the motion of each point is decomposed along fixed
directions. Moreover, this decomposition is usually carried out with respect
to two or three directions:\footnote{Depending on the dimension: Euler
treats both the two- and the three-dimensional cases.} the former is
appropriate for the decomposition, if the motion of all points is
completed in the same plane; but if their motion is not contained in
the same plane, it is appropriate to decompose the motion following
three fixed axes. Because the latter case is more difficult to treat,
it is more convenient to begin the investigation of possible motions
with the former case; once this has been done, the latter case will be
easily completed.

{\bf 10}.~~~First I will assign to the fluid two dimensions in such a
way that all of its particles are now not only found with certainty in
the same plane, but also their motion is performed in it. Let this
plane be represented in the plane of the table (Fig.~1), let an arbitrary point
$l$ of the fluid be considered, its position being denoted by
orthogonal coordinates $\mathrm{A} \mathrm{L} = x$ and $\mathrm{L}l =
y$.  The motion is decomposed following these directions, giving a
velocity $lm = u$ parallel to the axis $\mathrm{A} \mathrm{L}$ and $ln
= v$ parallel to the other axis $\mathrm{A} \mathrm{B}$: so that the
true future velocity of this point is $=\sqrt(uu + vv)$, and its
direction with respect to the axis $\mathrm{A} \mathrm{L}$ is inclined
by an angle with the tangent $\frac{v}{u}$.
\begin{figure}[!h]
   \includegraphics[scale=0.65]{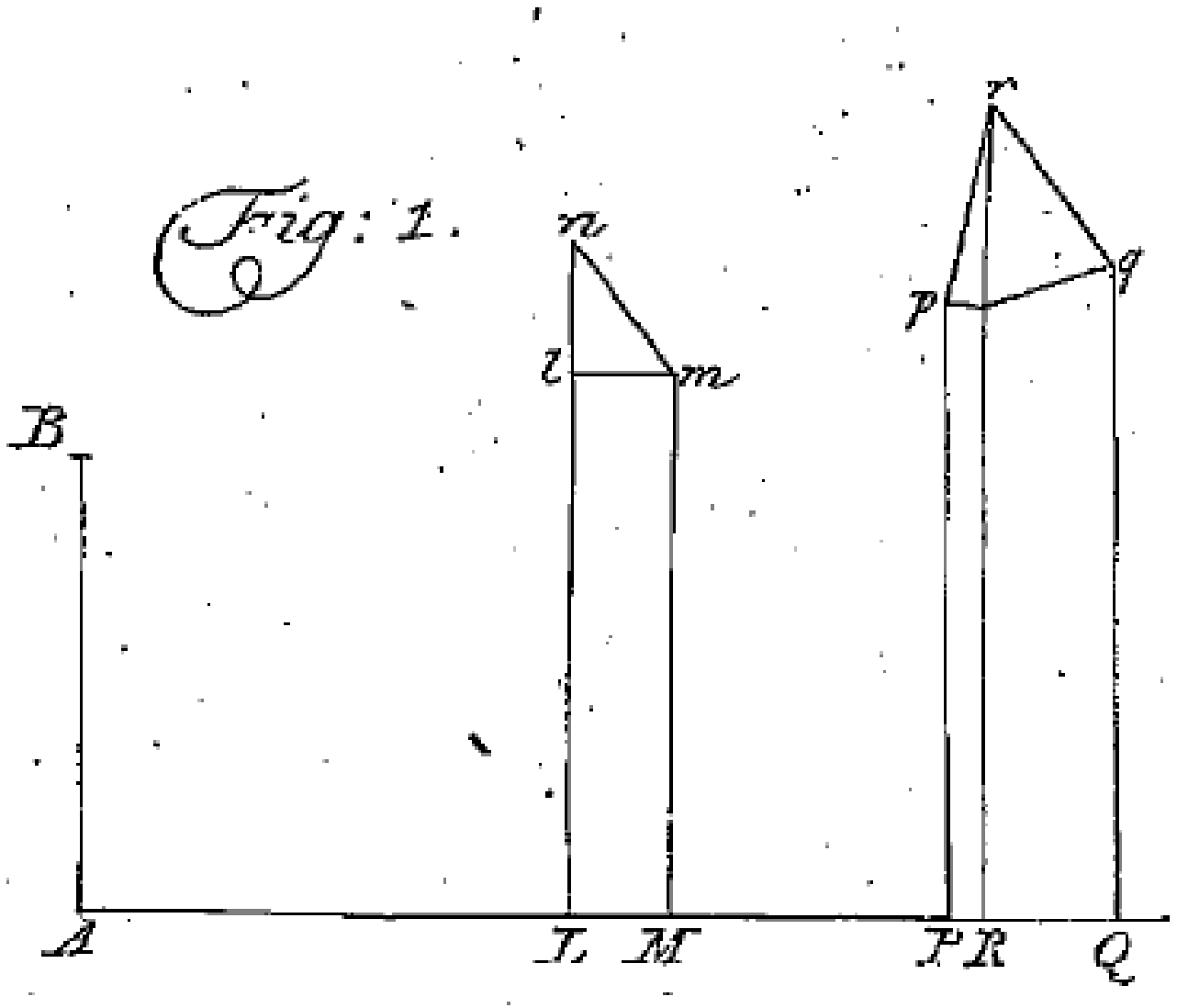}
\end{figure}

{\bf 11}.~~~Since the state of the motion, presented in a way which
suits the each point of the fluid, is supposed to evolve, the
velocities $u$ and $v$ will depend  on the position $l$ of the
point and will therefore be functions of the coordinates $x$ and
$y$. Thus, we put upon a differentiation
\begin{equation*}
du = \mathrm{L} dx + l dy \quad \mathrm{and} \quad 
dv = \mathrm{M} dx + m dy ,
\end{equation*}
which differential formulas, since they are complete,\footnote{Exact
differentials.} satisfy furthermore $\frac{d\mathrm{L} }{dy} =
\frac{dl }{dx} $ and $\frac{d\mathrm{M} }{dy} = \frac{d m}{dx} $.
Here it is to note that in such expression $\frac{d\mathrm{L}}{dy}$,
the differential of L itself or $d$L, is understood to be obtained
from the variability with respect to $y$; in similar manner in the
expression $dl/dx$, for $dl$ the differential of $l$ itself has to be
taken, which arises if we take  $x$ to vary.

{\bf 12}.~~~Thus, it is in order to be cautious and not to take in
such fractional expressions $\frac{d\mathrm{L}}{dy} $,
$\frac{dl}{dx}$, $\frac{d\mathrm{M}}{dy} $, and $\frac{dm}{dx} $ the
numerators $d$L, $dl$, $d$M, and $dm$ as denoting the complete
differentials of the functions $L$, $l$, $M$ and $m$; but constantly
they designate such differentials that are obtained from the
variation of only one coordinate, obviously the one, whose
differential is represented in the denominator; thus, such expressions
will always represent finite and well defined quantities.
Furthermore, in the same way are understood L$= \frac{du}{dx} $, $l =
\frac{du}{dy} $, M$= \frac{dv}{dx} $ and $m = \frac{dv}{dy} $; which
notation of ratios has been used for the first time by the most
enlightened \textit{Fontaine},\footnote{A
paper ``Sur le calcul int\'egral'' containing the notation $\frac{df}{dx}$ for the partial derivative of
$f$ with respect to $x$ was presented by Alexis Fontaine des Bertins to the
Paris Academy of Sciences in 1738, but it was published only a quarter of
a century later (Fontaine, 1764).  Nevertheless,
Fontaine's paper was widely known among mathematicians from the beginning
of the 1740s, and, particularly, was discussed in the correspondence between
Euler, Daniel Bernoulli and Clairaut; cf. Euler, 1980:~65--246.}
and  I will also apply it here, since it gives a non negligible
advantage  of calculation.

{\bf 13}.~~~Since $du = \mathrm{L} dx + l dy $ and $dv = \mathrm{M} dx
+ m dy $, here it is appropriate to assign a pair of velocities to the
point which is at an infinitely small distance from the point $l$; if
the distance of such a point from the point $l$ parallel to the axis
AL is $dx$, and parallel to the axis AB is $dy$, then the velocity of
this point parallel to the axis AL will be $u + \mathrm{L} dx + l dy
$; furthermore, the velocity parallel to the other axis AB is $ v +
\mathrm{M} dx + m dy $. Thus, during the infinitely short time $dt$
this point will be carried parallel to the direction of the axis AL
the distance $dt (u + \mathrm{L} dx + l dy )$ and parallel to the
direction of the other axis AB the distance $ dt ( v + \mathrm{M} dx
+ m dy )$.

{\bf 14}.~~~Having noted these things, let us consider a triangular
element $lmn$ of water, and let us seek the location into which it is
carried by the motion during the time $dt$. Let
$lm$ be the side parallel to the axis AL and let $ln$ be the side
parallel to the axis AB: let us also put $lm = dx$ and $ln = dy$; or
let the coordinates of the point $m$ be $x + dx $ and $y$; the
coordinates of the point $n$ be $x$ and $y + dy$. It is plain, since
we do not define the relation between the differentials $dx$ and $dy$,
which can be taken negative as well as positive, that in thought the
whole mass of fluid may be divided into elements of this sort, so that
what we determine for one in general will extend equally to all.

{\bf 15}.~~~To find out  how far the element $lmn$ is carried during the
time $dt$ due to the local motion, we search for the points $p$, $q$
and $r$, to which its vertices, or the points $l$, $m$ and $n$ are
transferred during the time $dt$. Since
$$
\begin{tabular}{c | c | c | c } 
& of point $l$ & of  point $m$ & of point $n$ \\
Velocity w.r.t.  AL$= $ & $u$ & $ u + \mathrm{L} dx $ & $ u + ldy $ \\
Velocity w.r.t. AB$= $  & $v$ & $ v + \mathrm{M} dx $ & $ v + mdy $ 
\end{tabular}
$$
in the time $dt $ the point  $l$  reaches the point $p$, chosen 
such that:
\begin{equation*}
\mathrm{A} \mathrm{P} - \mathrm{A} \mathrm{L} = u dt \quad \mathrm{and} \quad 
\mathrm{P} p - \mathrm{L}l = v dt .
\end{equation*}
Furthermore, the point $m$ reaches the point $q$, such that 
\begin{equation*}
\mathrm{A} \mathrm{Q} - \mathrm{A} \mathrm{M} = (u + \mathrm{L} dx )
dt \quad \mathrm{and} \quad \mathrm{Q} q - \mathrm{M} m = (v +
\mathrm{M} dx ) dt .
\end{equation*}
Also, the point $n$ is carried to $r$, chosen such that
\begin{equation*}
\mathrm{A} \mathrm{R} - \mathrm{A} \mathrm{L} = (u + l dy ) dt \quad \mathrm{and} \quad 
\mathrm{R} r - \mathrm{L} n = (v + mdy ) dt .
\end{equation*}

{\bf 16}.~~~Since the points $l$, $m$ and $n$ are carried to the
points $p$, $q$ and $r$, the triangle $lmn$ made of the joined
straight lines $pq$, $pr$ and $qr$, is thought to be arriving at the
location defined by the triangle $pqr$.  Because the triangle
$lmn$ is infinitely small, its sides cannot receive any curvature from
the motion, and therefore, after having performed the translation of
the element of water $lmn$ in the time $dt$, it will conserve the
straight and triangular form. Since this element $lmn$ must not
be either extended to a larger volume, nor compressed into a smaller
one, the motion should be arranged so that the volume of the triangle
$pqr$ is rendered to be equal to the area of the triangle $lmn$.

{\bf 17}.~~~The area of the triangle $lmn$, being rectangular
at $l$, is $=\frac{1}{2} dx dy $, value to which the area of the triangle
$pqr$ should be put equal. To find this area, the pair of coordinates of
the points $p$, $q$ and $r$ must be examined, which are:
\begin{equation*}
\begin{split}
& \mathrm{A} \mathrm{P} = x + u dt ; \quad \mathrm{A} \mathrm{Q} = x + dx + (u + \mathrm{L} dx ) dt ; \\
& \mathrm{A} \mathrm{R} = x + (u + ldy ) dt ;  \quad  \mathrm{P} p  = y + v dt \\
& \mathrm{Q} q = y + (v + \mathrm{M} dx) dt , \quad \mathrm{R} r = y + dy + (v + mdy ) dt 
\end{split}
\end{equation*}
Then, indeed, the area of the triangle $pqr$ is found from the area of
the succeeding trapezoids, so that
\begin{equation*}
pqr = \mathrm{P} pr \mathrm{R} + \mathrm{R} rq \mathrm{Q}  - \mathrm{P} pq \mathrm{Q} .
\end{equation*}
Since these trapezoids have a pair of sides parallel and perpendicular
to the base AQ, their areas are easily found.

{\bf 18}.~~~Plainly, these areas are given by the expressions
\begin{equation*}
\begin{split}
& \mathrm{P} pr \mathrm{R} = \frac{1}{2} \mathrm{P} \mathrm{R}
(\mathrm{P} p + \mathrm{R} r) \\ & \mathrm{R} rq \mathrm{Q} =
\frac{1}{2} \mathrm{R} \mathrm{Q} (\mathrm{R} r + \mathrm{Q} q) \\ &
\mathrm{P} pq \mathrm{Q} = \frac{1}{2} \mathrm{P} \mathrm{Q}
(\mathrm{P} p + \mathrm{Q} q)
\end{split}
\end{equation*}
By putting these together we  find:
\begin{equation*}
\Delta p q r = \frac{1}{2} \mathrm{P} \mathrm{Q} . \mathrm{R} r - \frac{1}{2} 
\mathrm{R} \mathrm{Q} . \mathrm{P} p - \frac{1}{2} \mathrm{P} \mathrm{R} . \mathrm{Q} q 
\end{equation*}
Let us  set for brevity
\begin{equation*}
\begin{split}
& \mathrm{A} \mathrm{Q} = \mathrm{A} \mathrm{P} + \mathrm{Q} ; \quad 
\mathrm{A} \mathrm{R} = \mathrm{A} \mathrm{P} + \mathrm{R} ; \quad 
\mathrm{Q} q = \mathrm{P} p + q ; \quad \mathrm{and} 
\quad \\
& \mathrm{R} r = \mathrm{P} p + r ,
\end{split}
\end{equation*}
so that $\mathrm{P}\mathrm{Q}=\mathrm{Q}$,
$\mathrm{P}\mathrm{R}=\mathrm{R}$, 
and $\mathrm{R}\mathrm{Q}=\mathrm{Q} - \mathrm{R}$, and we have $\Delta pqr  
= \frac{1}{2} 
\mathrm{Q} (\mathrm{P} p + r ) - \frac{1}{2} (\mathrm{Q} - \mathrm{R}) 
\mathrm{P} p - 
\frac{1}{2} \mathrm{R} (\mathrm{P} p + q) $ or
$\Delta pqr = \frac{1}{2} \mathrm{Q} . r - \frac{1}{2} \mathrm{R} . q$. 

{\bf 19}.~~~Truly, from the  values of the coordinates represented before 
it follows that
\begin{equation*}
\begin{split}
& \mathrm{Q} = dx + \mathrm{L} dx dt ; \quad q = \mathrm{M} dx dt \\
& \mathrm{R} =  l dy dt     ; \quad r = dy + m dy dt ,
\end{split}
\end{equation*}
upon the substitution of these values, the  area of the triangle is obtained 
\begin{equation*}
\begin{split}
& pqr = \frac{1}{2} dx dy ( 1 + \mathrm{L} dt ) ( 1 + m dt ) - \frac{1}{2} 
\mathrm{M} l \, dx dy dt^2 , \quad \mathrm{or} \\
& pqr = \frac{1}{2} dx dy ( 1 + \mathrm{L} dt + m dt + \mathrm{L} m dt ^2 - \mathrm{M} l dt^2 ) .
\end{split}
\end{equation*}
This should be equal to the area of the triangle $lmn$, that is
$=\frac{1}{2} dx dy $; hence we obtain the following equation 
\begin{equation*}
\begin{split}
& \mathrm{L} dt + m dt + \mathrm{L} m dt^2 - \mathrm{M} l dt^2 = 0
\quad \mathrm{or} \\ & \mathrm{L} + m + \mathrm{L} m dt - \mathrm{M} l
dt = 0 .
\end{split}
\end{equation*}

{\bf 20}.~~~~Since the terms L$mdt $ and M$l dt $ vanish for finite L
and $m$, we will have the equation $\mathrm{L} + m =0$. Hence, for the
motion to be possible, the velocities $u$ and $v$ of any point $l$
have to be arranged such that after calculating their differentials
\begin{equation*}
d u = \mathrm{L} dx + l dy , \quad \mathrm{and} \quad dv = \mathrm{M}
dx + m dy ,
\end{equation*}
one has $ \mathrm{L} + m = 0$. Or, since L$= \frac{du}{dx} $ and $m =
\frac{dv }{dy} $, the velocities $u$ and $v$, which are considered to
occur at the point $l$ parallel to the axes $AL$ and $AB$, must be
functions of the coordinates $x$ and $y$ such that $\frac{du}{dx} +
\frac{dv}{dy} = 0 $, and thus, the criterion of possible motions
consists in this that $\frac{du}{dx} + \frac{dv}{dy} = 0
$;\footnote{This is the two-dimensional incompressibility condition,
which in a slightly different form has already been established by
d'Alembert, 1752; cf. also Darrigol and Frisch, 2008:\S~III.} and unless
this condition holds, the motion of the fluid cannot take place.

{\bf 21}.~~~We shall proceed identically when the motion of the fluid
is not confined to the same plane.  Let us assume, to investigate this
question in the broadest sense, that all particles of the fluid are
agitated among themselves by an arbitrary motion, with the only law to
be respected that neither condensation nor expansion of the parts
occurs anywhere: in the same way, we seek which condition should apply
to the velocities that are considered to occur at every point, so that
the motion is possible: or, which amounts to the same, all motions
that are opposed to these conditions should be eliminated from the
possible ones, this being the criterion of possible motions.

{\bf 22}.~~~Let us consider an arbitrary point of the fluid $\lambda
$. To represent its location we use three fixed axes AL, AB and AC
orthogonal to each other (Fig.~2). Let the triple coordinates parallel to these
axes be AL$= x $, L$l = y$ and $l\lambda = z $; which are obtained if
firstly a perpendicular $\lambda l $ is dropped from the point
$\lambda $ to the plane determined by the two axes AL and AB; and then
a perpendicular $l$L is drawn from the point $l$ to the axis AL. In
this manner the location of the point $\lambda $ is expressed through
three such coordinates in the most general way and can be adapted to
all points of the fluid.
\begin{figure}[!h]
  \includegraphics[scale=0.33]{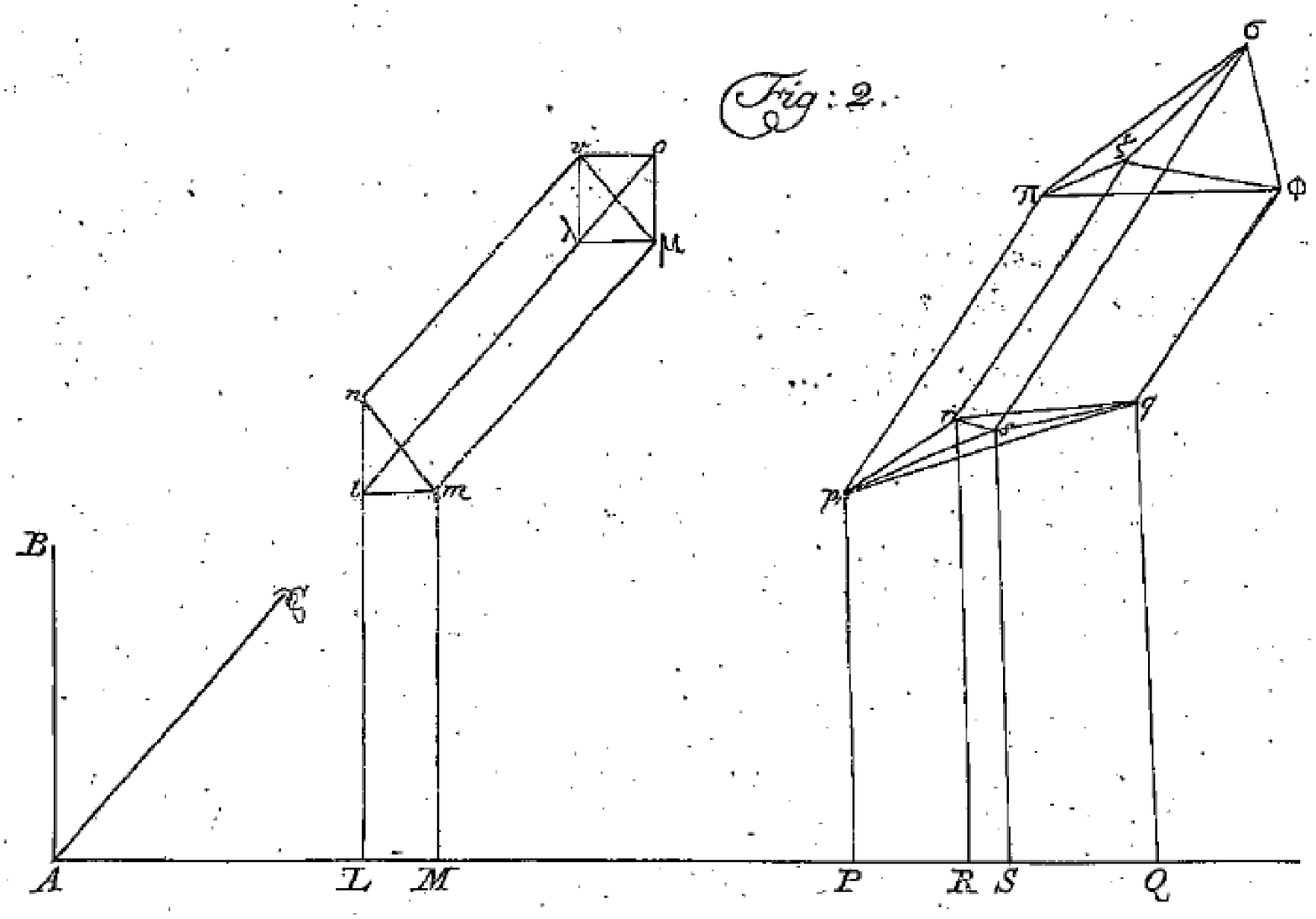}%
\end{figure}

{\bf 23}.~~~Whatever the later motion of the point $\lambda $, it can be
resolved following the three directions $\lambda \mu $, $\lambda \nu
$, $\lambda o $, parallel to the axes $AL$, $AB$ and $AC$. For the
motion of  the point $\lambda $ we set
\begin{equation*}
\begin{split}
& \mathrm{the\ velocity\ parallel\ to\ the\ direction} \quad \lambda \mu = u, \\
& \mathrm{the\ velocity\ parallel\ to\ the\ direction} \quad \lambda \nu = v, \\
& \mathrm{the\ velocity\ parallel\ to\ the\ direction} \quad \lambda o = w.   
\end{split}
\end{equation*}
Since these velocities can vary in an arbitrary manner for different
locations of the point $\lambda $, they will have to be considered as
functions of the three coordinates $x$, $y$ and $z$. After
differentiating them, let us put to proceed
\begin{equation*}
\begin{split}
& du = \mathrm{L} dx + l dy + \lambda dz \\
& dv = \mathrm{M} dx + m dy + \mu dz \\
& dw = \mathrm{N} dx + n dy + \nu dz .
\end{split}
\end{equation*}
Henceforth the quantities L, $l$, $\lambda $, M, $m$, $\mu $, N, $n$,
$\nu $ will be functions of the coordinates $x$, $y$ and $z$.

{\bf 24}.~~~ Because these formulas are complete differentials, we obtain
as above 
\begin{equation*}
\begin{split}
& \frac{d\makebox[1.6ex][l]{$\mathrm{L} $} }{dy } = 
\frac{d\makebox[1.6ex][l]{$l$}}{dx} ; 
\quad \frac{d\makebox[1.6ex][l]{$\mathrm{L} $} }{dz} 
= \frac{d\makebox[1.6ex][l]{$\lambda $} }{dx} ; \quad 
\frac{d\makebox[1.6ex][l]{$l$} }{dz} = \frac{d\makebox[1.6ex][l]{$\lambda $} }{dy} \\
& \frac{d\makebox[1.6ex][l]{$\mathrm{M} $} }{dy } = 
\frac{d\makebox[1.6ex][l]{$m$} }{dx} 
; \quad \frac{d\makebox[1.6ex][l]{$\mathrm{M} $} }{dz} 
= \frac{d\makebox[1.6ex][l]{$\mu $} }{dx} ; \quad 
\frac{d\makebox[1.6ex][l]{$m$} }{dz} = 
\frac{d\makebox[1.6ex][l]{$\mu $} }{dy} \\
& \frac{d\makebox[1.6ex][l]{$\mathrm{N} $}}{dy } = 
\frac{d\makebox[1.6ex][l]{$n$}}{dx} ; \quad 
\frac{d\makebox[1.6ex][l]{$\mathrm{N} $}}{dz} 
= \frac{d\makebox[1.6ex][l]{$\nu $}}{dx} ; \quad 
\frac{d\makebox[1.6ex][l]{$n$}}{dz} = 
\frac{d\makebox[1.6ex][l]{$\nu $}}{dy}, \\
\end{split}
\end{equation*}
where it is  assumed that the only varying coordinate is that  
whose differential appears in the denominator 
varies.\footnote{The partial differential
  notation was so new that Euler had to remind the reader of its definition.} 

{\bf 25}.~~~Thus, this point $\lambda $ will be moved in the time $dt$
by this threefold motion, which is considered to take place at the point X;
hence it moves
\begin{equation*}
\begin{split}
& \mathrm{parallel\ to\ the\ axis} \ \mathrm{AL}\  \mathrm{the\ distance} = u dt \\
& \mathrm{parallel\ to\ the\ axis} \ \mathrm{AB}\  \mathrm{the\ distance} = v dt \\
& \mathrm{parallel\ to\ the\ axis} \ \mathrm{AC}\  \mathrm{the\ distance} = w dt \\
\end{split}
\end{equation*}
The true velocity of the point $\lambda $, denoted by $=V$, which
clearly arises from the composition of this triple motion, is given in
view
 of orthogonality of the three directions by 
$V = \sqrt(uu + vv + ww )$ and the elementary distance, which is
traveled in the time $dt$ through its motion, will be $V dt$.

{\bf 26}.~~~Let us consider an arbitrary solid element of the fluid to
see whereto it is carried during the time $dt$; since it amounts
to the same, let us assign a quite arbitrary  shape to that element, but of 
a kind such  that the entire
mass of the fluid can be divided into such elements; to
investigate by calculation, let the shape be a right triangular pyramid,
bounded by four solid angles $\lambda $, $\mu $, $\nu $ and $o $, so
that for each one there are three coordinates
$$
\begin{tabular}{c | c | c | c | c } 
& of point $\lambda $ & of point $\mu $ & of point $\nu $ & of point $o$\\
w.r.t.  AL  & $x$ & $ x + dx $ & $ x $          & $x$     \\
w.r.t. AB  & $y$ & $ y $          & $ y + dy $ & $y$     \\
w.r.t. AC  & $z$ & $ z $          & $ z $          & $z + dz$
\end{tabular}
$$ Since the base of this pyramid is $\lambda \mu \nu = lmn =
\frac{1}{2} dx dy $ and the hight $\lambda o = dz$, its volume will be
$=\frac{1}{6} dx dy dz $.

{\bf 27}.~~~Let us investigate, whereto these vertices $\lambda $,
$\mu $, $\nu $ and $o$ are carried during the time $dt$: for which
purpose their three velocities parallel to the directions of the three
axes must be considered. The differential values of the
velocities $u$, $v$ and $w$ are given by
$$
\begin{tabular}{c | c | c | c | c } 
Velocity & of point $\lambda $ & of point $\mu $ & of point $\nu $ & of point $o$\\
w.r.t.  AL & $u$ & $ u + \mathrm{L} dx $   & $ u + l  dy  $   & $u + \lambda dz $     \\
w.r.t. AB  & $v$ & $ v + \mathrm{M} dx $   & $ v + m dy $  & $v + \mu dz$          \\
w.r.t. AC  & $w$ & $w + \mathrm{N} dx $  & $ w + n dy $   & $w + o dz$
\end{tabular}
$$

{\bf 28}.~~~If we let the points $\lambda $, $\mu $, $\nu $ and $o$ be transferred to the points $\pi $,
$\Phi $, $\rho $ and $\sigma $ in the time $dt$, and establish the
three coordinates of these points parallel to the axes, the small
displacement parallel to these axes will be 
\begin{center}
\begin{tabular}{r c l} 
\makebox[1.6ex][l]{A}\makebox[1.6ex][l]{P}$-$\makebox[1.6ex][l]{A}
\makebox[1.6ex][l]{L}  
& $= $& $u \, dt$         \\            
\makebox[1.6ex][l]{A}\makebox[1.6ex][l]{Q}$-$\makebox[1.6ex][l]{A}
\makebox[1.6ex][l]{M} & $= $& $(u+\makebox[1.6ex][l]{$\mathrm{L} $} \, dx) \, dt $  \\
\makebox[1.6ex][l]{A}\makebox[1.6ex][l]{R}$-$\makebox[1.6ex][l]{A}
\makebox[1.6ex][l]{L}  & $= $ &$(u + \makebox[1.6ex][l]{$l$} \, dy ) \, dt$    \\
\makebox[1.6ex][l]{A}\makebox[1.6ex][l]{S}$-$\makebox[1.6ex][l]{A}
\makebox[1.6ex][l]{L}   & $= $ &$(u + \makebox[1.6ex][l]{$\lambda $} \, dz ) \, dt$ \\
\hline
\makebox[1.6ex][l]{$\mathrm{P}$}\makebox[1.6ex][l]{$p$} $-$ 
\makebox[1.6ex][l]{$\mathrm{L}$}\makebox[1.6ex][l]{$l$} & $=$ & $ v \, dt$   \\
\makebox[1.6ex][l]{$\mathrm{Q}$}\makebox[1.6ex][l]{$q$} $-$ 
\makebox[1.6ex][l]{$\mathrm{M}$}\makebox[1.6ex][l]{$m$} & $=$ & $ 
(v + \makebox[1.7ex][l]{$\mathrm{M}$} \, dx) \, dt$  \\
\makebox[1.6ex][l]{$\mathrm{R} $}\makebox[1.6ex][l]{$r$} $-$ 
\makebox[1.6ex][l]{$\mathrm{L} $}\makebox[1.6ex][l]{$n$} & $=$ & $
(v + \makebox[1.6ex][l]{$m$} \, d y ) \, dt$      \\
\makebox[1.6ex][l]{$\mathrm{S} $}\makebox[1.6ex][l]{$s$} $-$ 
\makebox[1.6ex][l]{$\mathrm{L} $}\makebox[1.6ex][l]{$l$} &  $=$ & $
(v + \makebox[1.6ex][l]{$\mu $} \, dz ) \, dt$  \\
\hline
\makebox[1.6ex][l]{$p$}\makebox[1.6ex][l]{$\pi $}     $-$ 
\makebox[1.6ex][l]{$l$}\makebox[1.6ex][l]{$\lambda $} & $=$ & $ w \, dt $        \\
\makebox[1.6ex][l]{$q$}\makebox[1.6ex][l]{$\Phi $}    $-$ 
\makebox[1.6ex][l]{$m$}\makebox[1.6ex][l]{$\mu  $}    & $=$ & $ 
(w + \makebox[1.6ex][l]{$\mathrm{N} $} \, dx) \, dt $ \\ 
\makebox[1.6ex][l]{$r$}\makebox[1.6ex][l]{$\rho $}    $-$ 
\makebox[1.6ex][l]{$n$}\makebox[1.6ex][l]{$\nu  $}    & $=$ & $ 
(w + \makebox[1.6ex][l]{$n$} \, dy) \, dt $     \\
\makebox[1.6ex][l]{$s$}\makebox[1.6ex][l]{$\sigma $}  $-$ 
\makebox[1.6ex][l]{$l$}\makebox[1.6ex][l]{$o$}        & $=$ & $ 
(w + \makebox[1.6ex][l]{$\nu $} \, dz ) \, dt$ 
\end{tabular}  
\end{center}
Thus the three coordinates for these four points $\pi $, $\Phi $,
$\rho $ and $\sigma $ will be
\begin{eqnarray*}
\mathrm{AP} = x + u dt ; & & \mathrm{P} p = y + v dt ; \\
p \pi = z + w dt  & & \\
\mathrm{RQ} = x + dx + (u + \mathrm{L} dx) dt ; & & \mathrm{Q} q = y + (v + \mathrm{M} dx) dt ; \\
q \Phi = z + (w + \mathrm{N} dx ) dt &  & \\
\mathrm{AR} = x + (u + l dy) dt ; & & \mathrm{R} r = y + dy + (v + m dy) dt ; \\
r \rho = z + (w + n dy) dt  & & \\
\mathrm{AS} = x + (u + \lambda dz ) dt ; & & \mathrm{S} s = y + (v + \mu dz ) dt ; \\
s \sigma = z + dz + (w + \nu dz ) dt & &
\nonumber
\end{eqnarray*}

{\bf 29}.~~~Since after the time $dt$ has elapsed the vertices
$\lambda $, $\mu $, $\nu$ and $o$ of the pyramid are transferred to
the points $\pi $, $\Phi $, $\rho $ and $\sigma $, $\pi \Phi \rho
\sigma $ defines a similar triangular pyramid. Due to the nature of
the fluid the volume of the pyramid $\pi \Phi \rho \sigma $ should be
equal to the volume of the pyramid $\lambda \mu \nu o $ put forward,
that is $=\frac{1}{6} dx dy dz $.  Thus, the whole matter is reduced
to determining the volume of the pyramid $\pi \Phi \rho \sigma
$. Clearly, it remains a pyramid, if the solid $pqr \pi \Phi \rho
\sigma $ is removed from the solid $pqr \pi \Phi \rho \sigma $; the
latter solid is a prism orthogonally incident to the triangular basis
$pqr$, and cut by the upper oblique section $\pi \rho \Phi $.

{\bf 30}.~~~The other solid $p q r \pi \Phi \rho \sigma $ can be
divided by similarly into three prisms truncated in this manner, namely
\begin{equation*}
\mathrm{I}. \ pqrs \pi \Phi \sigma ; \quad \mathrm{II} . \ p r s \pi \rho \sigma ; \quad \mathrm{III} . \ 
q r s \Phi 
\rho \sigma 
\end{equation*} 
This has to be accomplished in such a way that
\begin{equation*}
\frac{1}{6} dx dy dz = p q s \pi \Phi \sigma + p r s \pi \rho \sigma + q r s \Phi \rho \sigma - 
p q r \pi \Phi \rho .
\end{equation*}
Since such a prism is orthogonally incident to its lower base, and
furthermore has three unequal heights, its volume is found by
multiplying the
base by one third of the sum of these heights.

{\bf 31}.~~~Thus, the volumes of these truncated prisms will be
\begin{equation*}
\begin{split}
& 
\makebox[1.6ex][l]{$p$}\makebox[1.6ex][l]{$q$}\makebox[1.6ex][l]{$s$} 
\makebox[1.6ex][l]{$\pi $}\makebox[1.6ex][l]{$\Phi $}\makebox[1.6ex][l]{$\sigma $}
= \frac{1}{3} \makebox[1.6ex][l]{$p$}\makebox[1.6ex][l]{$q$}
\makebox[1.6ex][l]{$s$} (\makebox[1.6ex][l]{$p$}\makebox[1.6ex][l]{$\pi $} + 
\makebox[1.6ex][l]{$q$}\makebox[1.6ex][l]{$\Phi $} + \makebox[1.6ex][l]{$s$} 
\makebox[1.6ex][l]{$\sigma $} ) \\
& 
\makebox[1.6ex][l]{$p$}\makebox[1.6ex][l]{$r$}\makebox[1.6ex][l]{$s$} 
\makebox[1.6ex][l]{$\pi $}\makebox[1.6ex][l]{$\rho $}\makebox[1.6ex][l]{$\sigma $}
= \frac{1}{3} \makebox[1.6ex][l]{$p$}\makebox[1.6ex][l]{$r$}\makebox[1.6ex][l]{$s$} 
(\makebox[1.6ex][l]{$p$}\makebox[1.6ex][l]{$\pi $} + \makebox[1.6ex][l]{$r$}
\makebox[1.6ex][l]{$\rho $} + \makebox[1.6ex][l]{$s$}\makebox[1.6ex][l]{$\sigma $} ) 
\\
& 
\makebox[1.6ex][l]{$q$}\makebox[1.6ex][l]{$r$}\makebox[1.6ex][l]{$s$} 
\makebox[1.6ex][l]{$\Phi $}\makebox[1.6ex][l]{$\rho $}\makebox[1.6ex][l]{$\sigma $}
= \frac{1}{3} \makebox[1.6ex][l]{$q$}\makebox[1.6ex][l]{$r$}\makebox[1.6ex][l]{$s$} 
(\makebox[1.6ex][l]{$q$}\makebox[1.6ex][l]{$\Phi $} + \makebox[1.6ex][l]{$r$} 
\makebox[1.6ex][l]{$\rho $} + \makebox[1.6ex][l]{$s$}\makebox[1.6ex][l]{$\sigma $} ) 
\\
& 
\makebox[1.6ex][l]{$p$}\makebox[1.6ex][l]{$q$}\makebox[1.6ex][l]{$r$} 
\makebox[1.6ex][l]{$\pi $}\makebox[1.6ex][l]{$\Phi $}\makebox[1.6ex][l]{$\rho $}
= \frac{1}{3} \makebox[1.6ex][l]{$p$}\makebox[1.6ex][l]{$q$}\makebox[1.6ex][l]{$r$} 
(\makebox[1.6ex][l]{$p$}\makebox[1.6ex][l]{$\pi $} + \makebox[1.6ex][l]{$q$} 
\makebox[1.6ex][l]{$\Phi $} + \makebox[1.6ex][l]{$r$}\makebox[1.6ex][l]{$\rho $} ) .
\end{split}
\end{equation*}
Since $pqr = pqs + prs + qrs $, the sum of the first three prisms will
definitely be small, or
\begin{equation*}
\frac{1}{6} dx dy dz = - \frac{1}{3} p \pi . qrs - \frac{1}{3} q \Phi . prs - \frac{1}{3} r \rho . 
pqs + \frac{1}{3} s \sigma . pqr ,
\end{equation*}
or
\begin{equation*}
dx dy dz = 2 pqr . s \sigma - 2 pqs . r \rho - 2 prs . q \Phi - 2 qrs . p \pi .
\end{equation*}

{\bf 32}.~~~Thus, it remains to define the bases of these prisms: but before we do this, let us put
\begin{equation*}
\begin{split}
& 
\makebox[3.2ex][l]{$\mathrm{AQ}$} = \makebox[3.2ex][l]{$\mathrm{AP}$} + 
\makebox[1.6ex][l]{$\mathrm{Q}$} ; \quad 
\makebox[3.2ex][l]{$\mathrm{Q}q$} = \makebox[3.2ex][l]{$\mathrm{P} p$} + 
\makebox[1.6ex][l]{$q$} ; 
\quad \makebox[3.2ex][l]{$q\Phi $} = \makebox[3.2ex][l]{$p \pi $} + 
\makebox[1.6ex][l]{$\Phi $} ; \\
& \makebox[3.2ex][l]{$\mathrm{AR}$} = \makebox[3.2ex][l]{$\mathrm{AP}$} + 
\makebox[1.6ex][l]{$\mathrm{R}$} ; \quad 
\makebox[3.2ex][l]{$\mathrm{R} r$} = \makebox[3.2ex][l]{$\mathrm{P} p$} + 
\makebox[1.6ex][l]{$r$} ; 
\quad 
\makebox[3.2ex][l]{$r \rho $}= \makebox[3.2ex][l]{$p \pi $}+ 
\makebox[1.6ex][l]{$\rho $} ; \\
& 
\makebox[3.2ex][l]{$\mathrm{AS}$} = \makebox[3.2ex][l]{$\mathrm{AP}$} + 
\makebox[1.6ex][l]{$\mathrm{S}$} ; \quad 
\makebox[3.2ex][l]{$\mathrm{S} s$} = \makebox[3.2ex][l]{$\mathrm{P} p$} + 
\makebox[1.6ex][l]{$s$} ; 
\quad 
\makebox[3.2ex][l]{$s \sigma $} = \makebox[3.2ex][l]{$p \pi $} + 
\makebox[1.6ex][l]{$\sigma $} , \\
\end{split}
\end{equation*} 
in order to shorten the following calculations. After the
substitution of these values, the terms containing $p \pi $ will
annihilate each other, and we shall have
\begin{equation*}
dx dy dz = 2 pqr . \sigma - 2 pqs . \rho - 2 prs . \Phi 
\end{equation*}
so that the value of the bases to be investigated is smaller. 

{\bf 33}.~~~Furthermore the triangle $pqr$ is obtained by removing 
the trapezoid
P$pq$Q from the figure P$prq$Q, the latter being the sum of the
trapezoids P$pr\mathrm{R}$ and $\mathrm{R} rq$Q; from which  it follows that
\begin{equation*}
\Delta pqr = \frac{1}{2} \mathrm{PR} (\mathrm{P} p + \mathrm{R} r) +
\frac{1}{2} \mathrm{RQ} (\mathrm{R} r + \mathrm{Q} q) - \frac{1}{2}
\mathrm{PQ} (\mathrm{P}p + \mathrm{Q}q) ;
\end{equation*}
or, because of $\mathrm{P}\mathrm{R}=\mathrm{R}$; $\mathrm{R}\mathrm{Q}
=\mathrm{Q}-\mathrm{R}$; and $\mathrm{P}\mathrm{Q} = \mathrm{Q}$ we
shall have
\begin{equation*}
\Delta  p q r =  \frac{1}{2} \mathrm{R} (\mathrm{P} p - \mathrm{Q} q ) + 
\frac{1}{2} \mathrm{Q} (\mathrm{R} r - \mathrm{P} p) = \frac{1}{2} \mathrm{Q} r - \frac{1}{2} \mathrm{R} q .
\end{equation*}
In the same manner we have
\begin{equation*}
\begin{split}
& \Delta p q s = \frac{1}{2} \mathrm{PS} (\mathrm{P} p + \mathrm{S} s) + 
\frac{1}{2} \mathrm{SQ} (\mathrm{S} s + \mathrm{Q} q) - \frac{1}{2} \mathrm{PQ} 
(\mathrm{P} p + \mathrm{Q} q) , \\
& \mathrm{or} \\
& \Delta p q s = \frac{1}{2} \mathrm{S} (\mathrm{P} p + \mathrm{S} s) + 
\frac{1}{2} (\mathrm{Q} - \mathrm{S} ) (\mathrm{S} s + \mathrm{Q} q) - \frac{1}{2} \mathrm{Q} 
(\mathrm{P} p + \mathrm{Q} q) , \\
\end{split}
\end{equation*}
from where it follows that:
\begin{equation*}
\Delta p q s = \frac{1}{2} \mathrm{S} (\mathrm{P} p - \mathrm{Q} q) + \frac{1}{2} \mathrm{Q} 
(\mathrm{S} s - \mathrm{P} p) = \frac{1}{2} \mathrm{Q} s - \frac{1}{2} \mathrm{S} q .
\end{equation*}
And finally
\begin{equation*}
\begin{split}
& \Delta p r s = \frac{1}{2} \mathrm{PR} (\mathrm{P} p + \mathrm{R} r) + 
\frac{1}{2} \mathrm{RS} (\mathrm{R} r + \mathrm{S} s) - \frac{1}{2} 
\mathrm{PS} (\mathrm{P} p + \mathrm{S} s) , \\
& \mathrm{or} \\
&  \Delta p r s = \frac{1}{2}  \mathrm{R} (\mathrm{P} p + \mathrm{R} r) + 
\frac{1}{2} (\mathrm{S} - \mathrm{R} ) (\mathrm{R} r + \mathrm{S} s) - \frac{1}{2} 
\mathrm{S} (\mathrm{P} p + \mathrm{S} s) 
\end{split}
\end{equation*}
from where it follows that
\begin{equation*}
\Delta p r s = \frac{1}{2} \mathrm{R} (\mathrm{P} p - \mathrm{S} s) + \frac{1}{2} \mathrm{S} 
(\mathrm{R} r - \mathrm{P} p) = \frac{1}{2} \mathrm{S} r - \frac{1}{2} \mathrm{R} s.
\end{equation*}

{\bf 34}.~~~After the substitution of these values we will obtain
\begin{equation*}
dx dy dz = (\mathrm{Q} r - \mathrm{R} q) \sigma + (\mathrm{S} q - \mathrm{Q} s) \rho + 
(\mathrm{R} s - \mathrm{S} r) \Phi; 
\end{equation*}
thus  the volume of the pyramid $\pi \Phi \rho \sigma $  will be 
\begin{equation*}
\frac{1}{6} (\mathrm{Q} r - \mathrm{R} q) \sigma + \frac{1}{6} (\mathrm{S} q - \mathrm{Q} s) \rho + 
\frac{1}{6} (\mathrm{R} s - \mathrm{S} r) \Phi . 
\end{equation*}
From the values of the coordinates presented above in $\S.~28$ follows
\begin{eqnarray*}
\mathrm{Q} = dx + \mathrm{L} dx dt    &  q = \mathrm{M} dx dt          &  
\Phi = \mathrm{N} dx dt  \nonumber \\
\mathrm{R} = l dy dt              &  r  = dy + m dy dt  & \rho = n dy dt                        \\
\mathrm{S} = \lambda dz dt & s = \mu dz dt         & \sigma = dz + \nu dz dt .  \nonumber   
\end{eqnarray*}

{\bf 35}.~~~Since here we have
\begin{equation*}
\begin{split}
& 
\makebox[3.2ex][l]{$\mathrm{Q} r$}  - \makebox[3.2ex][l]{$\mathrm{R}q$} 
= dx dy (1 + \mathrm{L} dt + m dt + \mathrm{L} m dt^2 - 
\mathrm{M} l dt^2 )            \\
& 
\makebox[3.2ex][l]{$\mathrm{S} q$}  - \makebox[3.2ex][l]{$\mathrm{Q} s$} 
= dx dz ( - \mu dt - \mathrm{L} \mu dt^2  + \mathrm{M} 
\lambda dt^2 )  \\
& 
\makebox[3.2ex][l]{$\mathrm{R} s$} - \makebox[3.2ex][l]{$\mathrm{S} r$}    
= dy dz ( - \lambda dt - m \lambda dt^2 + l \mu dt^2 ) 
\end{split}
\end{equation*}
the volume of the  pyramid $\pi \Phi \rho \sigma $ is found to be expressed as
\begin{equation*}
\frac{1}{6} dx dy dz \left\{ 
\begin{array}{l l l l}
1  
& 
+ \makebox[1.6ex][l]{$\mathrm{L}$} \, dt   
&  
+ \makebox[3.2ex][l]{$\mathrm{L} m$} \, dt^2  
& 
+ \makebox[4.8ex][l]{$\mathrm{L} m \nu $} \, dt^3 
\\
& 
+ \makebox[1.6ex][l]{$m$} \, dt   
&  
- \makebox[3.2ex][l]{$\mathrm{M} l$} \, dt^2   
& 
- \makebox[4.8ex][l]{$\mathrm{M} l \nu $} \, dt^3 
\\
& 
+ \makebox[1.6ex][l]{$\nu $} \,  dt 
& 
+ \makebox[3.2ex][l]{$\mathrm{L} \nu $} \, dt^2 
& 
- \makebox[4.8ex][l]{$\mathrm{L} n \mu $} \, dt^3 
\\
&               
& 
+ \makebox[3.2ex][l]{$m \nu $} \, dt^2   
& 
+ \makebox[4.8ex][l]{$\mathrm{M} n \lambda $} \, dt^3 
\\
&               
& 
- \makebox[3.2ex][l]{$n \mu $} \, dt^2 
& 
- \makebox[4.8ex][l]{$\mathrm{N} m \lambda $} \, dt^3 
\\
&               
& 
- \makebox[3.2ex][l]{$\mathrm{N} \lambda $} \, dt^2 
& 
+ \makebox[4.8ex][l]{$\mathrm{N} l \mu $} \, dt^3  
\end{array} \right\}  ,
\end{equation*}
which (volume), since it must be equal to that of the pyramid $\lambda
\mu \nu o = \frac{1}{6} dx dy dz $, will satisfy, after performing a
division by $dt $ the following  equation\footnote{This is the calculation to
which Euler refers in his later French memoir Euler, 1755.}
\begin{equation*}
\begin{split}
0 =  & \mathrm{L} + m + \nu + dt ( \mathrm{L} m + \mathrm{L} \nu + m \nu - \mathrm{M} l - 
\mathrm{N} \lambda - n \mu )    \\
        & + dt^2 ( \mathrm{L} m \nu + \mathrm{M} n \lambda + \mathrm{N} l \mu  - 
        \mathrm{L} n \mu - \mathrm{M} l \nu - \mathrm{N} l \mu ) .
\end{split}
\end{equation*}

{\bf 36}.~~~Discarding the infinitely small terms, we get this
equation:\footnote{This is the three-dimensional incompressibility
condition.} L$+ m + \nu = 0$, through which is determined the
relation between $u$, $v$ and $w$, so that the motion of the fluid be
possible. Since L$= \frac{du}{dx} $, $m = \frac{dv}{dy} $ and $\nu =
\frac{dw}{dz} $, at an arbitrary
point of the fluid $\lambda $, whose position is defined by the three
coordinates $x$, $y$ and $z$, and the velocities $u$,
$v$ and $w$ are assigned in the same manner to be directed along these same 
coordinates, the criterion of possible motions is such that
\begin{equation*}
\frac{du}{dx} + \frac{dv}{dy} + \frac{dw}{dz} = 0.
\end{equation*}  
This condition expresses that through the motion no part of the
fluid is carried into a greater or or lesser space, but perpetually
the continuity of the fluid as well as the identical density is conserved.

{\bf 37}.~~~This property is to be interpreted further so that at the
same instant it is extended to all points of the fluid:
of course, the three velocities of all the points must be
such functions of the three coordinates $x$, $y$ and $z$ that we have
$\frac{du}{dx} + \frac{dv}{dy} + \frac{dw}{dz} = 0$: in this way the
nature of those functions defines the motion of every point of the
fluid at a given instant. At another time the motion of the same
points may be howsoever different, provided that at an arbitrary point
of time the  property holds for the whole fluid. Up to now I
have considered the time simply as a constant quantity.

{\bf 38}.~~~If however, we also wish to consider the time as variable
so that the motion of the point $\lambda $ whose position is given by
the three coordinates AL$= x$, L$l = y $ and $l\lambda = z$ has to be
defined after the elapsed time $t$, it is certain that the three
velocities $u$, $v$ and $w$ depend not only on the coordinates $x$,
$y$ and $z$ but additionally on the time $t$, that is they will be
functions of these four quantities $x$, $y$, $z$ and $t$; furthermore,
their differentials are going to have the following form
\begin{equation*}
\begin{split}
& d u = \mathrm{L} dx +  l   dy + \lambda dz + \mathfrak{L} dt ;  \\
& d v = \mathrm{M} dx + m dy +  \mu dz + \mathfrak{M} dt       ;  \\
& d w = \mathrm{N} dx + n dy +  \nu dz  + \mathfrak{N} dt       ;  
\end{split}
\end{equation*}
Meanwhile we shall always have $\mathrm{L}+ m + \nu = 0 $, therefore at
every arbitrary instant the time $t$ is considered to be constant, or
$dt = 0$. Howsoever the functions $u$, $v$ and $w$ vary with time $t$,
it is necessary that at every moment of time the following holds:
\begin{equation*}
\frac{du}{dx} + \frac{dv}{dy} + \frac{dw}{dz} = 0 .               
 \end{equation*}
Since the condition expresses that any arbitrary portion of the
fluid is carried in a time $dt$ into a volume equal to itself, the
same will have to happen, due to the same condition, in the next time
interval, and therefore in all the following time intervals.

\section{Second part}
\label{s:parsaltera}

{\bf 39}.~~~Having exposed what pertains to all possible motions, let
us now investigate the nature of the motion which can really occur in
the fluid. Here, besides the continuity of the fluid and the constancy
of its density, we will also have to  consider the forces which act
on every element of the fluid. When the motion of any element is
either non-uniform or varying in its direction, the change of the
motion must be in accordance with the forces acting on this
element.  The change of the motion becomes known from
the known forces, and the preceding formulas contain this change; we
will now deduce  new conditions\footnote{Here Euler probably has in mind 
the condition of potentiality, which he will obtain in $\S\S.~47$ and 54 for 
the two-dimensional case and in $\S.~60$ for the three-dimensional case.}
which single out the actual motion
among all those possible up to this point.

{\bf 40}.~~~Let us arrange this investigation in two parts as well; at
first let us consider all motions being performed in the same
plane. Let $\mathrm{A}\mathrm{L}= x$, $\mathrm{L}l = y$ be, as before, the defining coordinates
of the position of an arbitrary point $l$; now, after the elapsed time
$t$, the two velocities of the point $l$ parallel to the axes AL and
AB are $u$ and $v$: since the variability of time has to be taken into
account, $u$ and $v$ will be functions of $x$, $y$ and $t$ themselves.
in respect of which we put
\begin{equation*}
d u = \mathrm{L} dx + l dy + \mathfrak{L} dt \quad \mathrm{and} 
d v = \mathrm{M} d x + m d y + \mathfrak{M} dt 
\end{equation*}
and we have established above that because of the former condition
encoutered above, we have $\mathrm{L}+ m = 0$.

{\bf 41}.~~~After an elapsed small time interval $dt$
the point $l$ is carried to $p$, and it has travelled a distance $u
dt$ parallel to the axis AL, a distance $v dt$ parallel to the other
axis AB. Hence, to obtain the increments in the velocities $u$ and
$v$ of the point $l$ which are induced during the time $dt$, for $dx$
and $dy$ we must write the distance $u dt$ and $v dt $, from which  will
arise these true increments of the velocities
\begin{equation*}
d u = \mathrm{L} u dt + l v dt + \mathfrak{L} dt \quad \mathrm{and} \quad 
d v = \mathrm{M} u dt + m v dt + \mathfrak{M} dt. 
\end{equation*}
Therefore the accelerating forces, which produce these accelerations
are
\begin{equation*}
\begin{split}
&  \mathrm{Accel.\ force\ w.r.t.} \ \mathrm{AL} = 2(\mathrm{L}u + lv
  +\mathfrak{L}) \\
&  \mathrm{Accel.\ force\ w.r.t.} \ \mathrm{AB} = 2(\mathrm{M}u + mv +\mathfrak{M})   
\end{split}
\end{equation*}
to which therefore the forces acting upon the particle of
water ought to be equal.\footnote{The unusual factors of 2 in the
  previous
equations have to do with a choice of units which soon became
obsolete; cf. Truesdell, 1954; Mikahailov, 1999.}

{\bf 42}.~~~Among the forces which in fact act upon
the particles of water, the first to be cñonsidered is gravity;
its effect, however, if the plane of motion is horizontal, amounts to
nothing. Yet if the plane is inclined, the axis AL
following the inclination, the other being horizontal,  gravity generates
a constant accelerating force parallel to the axis AL, let it be
$\alpha $.  Next we must not neglect friction, which often hinders
the motion of water, and not a little. Although its laws have not yet been
explored sufficiently, nevertheless, following the law of friction for solid
bodies, probably we shall not wander too far astray if we set the
friction everywhere proportional to the pressure with which the
particles of water press upon one another.\footnote{It is actually not
clear why Euler takes the friction force proportional to the
pressure.}

{\bf 43}.~~~First, must be brought into the calculation the pressure
with which the particles of water everywhere mutually act upon each
other, by means of which every particle is pressed together on all
sides by its neighbors; and in so far as this pressure is not
everywhere equal, to that extent motion is communicated to that
particle.\footnote{Here Euler makes full use of the concept of
internal pressure, cf. Darrigol and Frisch, 2008.} The water simply
will be put everywhere into a state of compression similar to
that which still water experiences when stagnating at a certain depth.
This depth is most conveniently employed for representing the
pressure at an arbitrary point $l$ of the fluid. Therefore let that
height, or depth, expressing the state of compression at $l$, be $p$,
a certain function of the coordinates $x$ and $y$, and should the pressure
at $l$ vary also with the time, the time  will also enter into the
function $p$.

{\bf 44}.~~~Thus let us set $dp = \mathrm{R} dx + r dy + \mathfrak{R}
dt $, and let us consider a rectangular element of water, $lmno$,
whose sides are $lm = no = dx $ and $ln = mo = dy $, whose area is $dx
dy$ (Fig.~3). The pressure at $l$ is $p$, the pressure at $m$ is  $ p +
\mathrm{R} dx $, at $ n$  it is $ p + r dy $ and at $o$ it is $ p + \mathrm{R} dx + r
dy $. Thus the side $lm$ is pressed by a force $ = dx (p + \frac{1}{2}
\mathrm{R} dx ) $, while the opposite side $no$ will be pressed by a
force $ = dx ( p + \frac{1}{2} \mathrm{R} dx + r dy ) $; therefore by
these two forces the element $lmno$ will be impelled in the direction
$ln$ by a force $ = - r dx dy $. Moreover, in a similar manner from
the forces $dy (p + \frac{1}{2} r dy )$ and $dy ( p + \mathrm{R} dx +
\frac{1}{2} r dy )$, which act on the sides $ln$ and $mo$ will result
a force $ = - \mathrm{R} dx dy $ impelling the element in the
direction $lm$.
\begin{figure}[!h]
\includegraphics[scale=0.8]{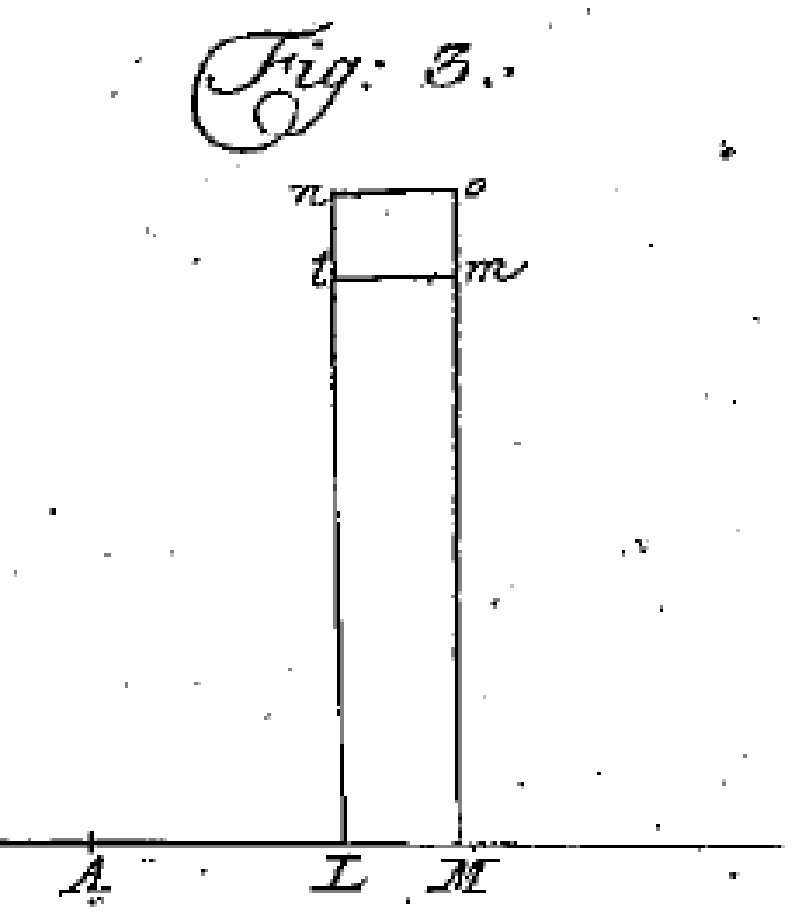}%
\end{figure}

{\bf 45}.~~~Thus will originate an accelerating force parallel to
$lm= - \mathrm{R}$ and an accelerating force parallel to $ln = - r$,
of which the one directed along the force of gravity $\alpha $
gives $\alpha - \mathrm{R}$. Having ignored  friction so far, we
obtain the following equations:\footnote{Here the so-called Euler
  equations of incompressible fluid dynamics appear for the first
  time, but the notation and the units are not very modern, in
  contrast to the memoir he will write three years later (Euler, 1755).}
\begin{equation*}
\begin{split}
\alpha 
& - \mathrm{R} = 2 L u + 2 l v + 2 \mathfrak{L} \quad \mathrm{or} \quad 
\mathrm{R} = \alpha - 2 \mathrm{L} u - 2 l v - 2 \mathfrak{L}   \\
& - r = 2 \mathrm{M} u + 2 m v + 2 \mathfrak{M}  \quad \mathrm{and} \quad 
r = - 2 \mathrm{M} u - 2 m v - 2 \mathfrak{M} 
\end{split}
\end{equation*}
from which we gather that 
\begin{equation*}
dp = \alpha dx - 2 (\mathrm{L} u + l v + \mathfrak{L} ) d x - 2 (\mathrm{M} u + m v + \mathfrak{M} ) d y + \mathfrak{R}  dt,
\end{equation*}
a differential which must be complete or integrable.

{\bf 46}.~~~Because the term $\alpha \, dx $ is integrable by itself
and nothing is determined for $\mathfrak{R} $, from the nature of
complete differentials it is necessary that the following holds in the notation
already employed:
\begin{equation*}
\frac{ d. \mathrm{L} u + l v + \mathfrak{L} }{dy } = \frac{
d. \mathrm{M} u + m v + \mathfrak{M} }{dx}.
\end{equation*}
Since $\frac{du}{dx} = $L, $\frac{du}{dy} = l $; $\frac{dv}{dx} = $M, and $\frac{dv}{dy} = m $  
it follows that
\begin{equation*}
\mathrm{L}  l + \frac{ u d\mathrm{L} }{dy} + l m + \frac{v d l}{dy} + \frac{d\mathfrak{L} }{dy} = 
\mathrm{M} \mathrm{L} + \frac{u d\mathrm{M} }{dx} +
m \mathrm{M} + \frac{v dm}{dx} = \frac{d \mathfrak{M} }{dx} 
\end{equation*}
which is reduced to this form:
\begin{equation*}
\begin{split}
& (\mathrm{L} + m) (l - \mathrm{M})  + \\
& u \left( \frac{d\mathrm{L} }{dy} - \frac{d\mathrm{M} }{dx} \right) + 
v \left( \frac{dl}{dy} - \frac{dm}{dx} 
\right) + \frac{d\mathfrak{L} }{dy}  - \frac{\mathfrak{M} }{dx}  = 0
\end{split}
\end{equation*}

{\bf 47}.~~~In fact, since we knew $\mathrm{L} d x + l dy + \mathfrak{L} dt $ and $\mathrm{M} dx + m dy + 
\mathfrak{M} dt $ to be complete differentials, 
\begin{equation*}
\frac{d\mathrm{L} }{dy} = \frac{dl}{dx} ;  \quad \frac{dm}{dx} = \frac{d\mathrm{M} }{dy} ; \quad 
\frac{d\mathfrak{L} }{dy } = \frac{d l}{dt} \quad \mathrm{and} \quad 
\frac{\mathfrak{M} }{dx} = \frac{d\mathrm{M} }{dt} 
\end{equation*}
after the substitution of which values we have the following equation
\begin{equation*}
\begin{split}
& (\mathrm{L} + m)(l - \mathrm{M} ) + \\
& u \left( \frac{dl - d\mathrm{M} }{dx} \right) + v \left( \frac{dl - d\mathrm{M} }{dy} \right) + 
\frac{dl - d\mathrm{M} }{dt} = 0. 
\end{split}
\end{equation*}
Plainly, this is satisfied if $l = \mathrm{M}$: so that $\frac{du}{dy}
= \frac{dv}{dx} $. Since this condition requires
that $\frac{du}{dy} = \frac{dv}{dx} $,\footnote{Here there are two
  problems. The minor problem is a typographical error in the
  published version ($\frac{du}{dx}$ instead of $\frac{dv}{dx}$), which is not present in a 1752 copy of the manuscript (not in
  Euler's hand), henceforth referred to as Euler, 1752. A more serious
  problem is that Euler here repeats the mistake of d'Alembert, 1752
  who confused a sufficient condition -- the vanishing of the
  vorticity -- with a necessary one.}  it appears
finally that the differential formula $u dx + v dy $ must be
complete;  in this lies the criterion of actual motion.

{\bf 48}.~~~This criterion is independent from the preceding one,
which was provided by the continuity of the fluid and its uniform
constant density. Therefore even if the fluid in motion changes its
density, as happens in the motion of elastic fluids such as air, this
property will hold nonetheless, namely $u dx + v dy $ has to  be a
complete differential. In other words, the velocities $u$ and $v$
must always be functions of the coordinates $x$ and $y$, together with
time $t$, in such a way that when  the time is taken constant the
formula $u dx + v dy $ admits an integration.

{\bf 49}.~~~We shall  now determine the pressure $p$ itself, which is
absolutely necessary for perfectly determining the motion of the
fluid. Since we have found that $\mathrm{M} = l$ we have
\begin{equation*}
\begin{split}
dp = \alpha dx - 2 u (\mathrm{L} dx + l dy) - 2 v (l dx + m dy)  & - 2 \mathfrak{L} dx  \\
 - 2 \mathfrak{M} dy  & + \mathfrak{R} dt. 
\end{split}
\end{equation*}
Moreover $\mathrm{L} dx + l dy = du - \mathfrak{L} dt $; $l dx + m dy =
dv - \mathfrak{M} dt $; hence we have
\begin{equation*}
\begin{split}
& dp = \alpha dx  - \\
& 2 u du - 2 v dv + 2 \mathfrak{L} u dt  + 2 \mathfrak{M} v dt - 2 \mathfrak{L} dx -
2 \mathfrak{M} dy + \mathfrak{R} dt 
\end{split}
\end{equation*}
Therefore, if we wish to ascertain for the present time the
pressure at each point of the fluid, with no account of its
variation in time, we shall have to consider this equation
\begin{equation*}
dp = \alpha dx - 2 u du - 2 v dv - 2 \mathfrak{L} dx - 2 \mathfrak{M} dy  ,
\end{equation*}
and in our notation $\mathfrak{L} = \frac{du}{dt} $ and $\mathfrak{M}
= \frac{dv}{dt} $.\footnote{The printed version has $L =
  \frac{dv}{dt}$ instead of $\mathfrak{L} = \frac{du}{dt}$. Euler, 1752 is correct.} Hence
\begin{equation*}
dp = \alpha dx - 2 u du - 2 v dv - 2 \frac{du}{dt} dx - 2 \frac{dv}{dt} dy  ,
\end{equation*}
in the integration of which the time is to be taken constant.

{\bf 50}.~~~This equation is integrable by hypothesis, and
is indeed understood as such, if we consider the criterion of the
motion  which, as we  have seen, consists in that $u dx + v dy$ be a
complete differential when the time $t$ is taken constant. Let 
therefore S be its integral, which consequently will be a function of
$x$, $y$ and $t$ themselves. For  $dt = 0$ we obtain
$d\mathrm{S} = u dx + v dy $, while assuming the time $t$
variable as well, let us write
\begin{equation*}
d\mathrm{S} = u dx + v dy + \mathrm{U} dt ,
\end{equation*}
on which account we obtain $\frac{du}{dt} = \frac{d\mathrm{U} }{dx} $ 
and $\frac{dv}{dt} = 
\frac{d\mathrm{U} }{dy} $. 
Then, in fact $\mathrm{U} = \frac{d\mathrm{S} }{dt} $. 

{\bf 51}.~~~After inserting these values we will obtain
\begin{equation*}
\frac{du}{dt} . dx + \frac{dv}{dt} . dy = \frac{d\mathrm{U} }{dx} . dx + \frac{d\mathrm{U} }{dy} . dy 
\end{equation*}
and this differential formula is manifestly integrated at constant
time $t$ to give $\mathrm{U} $. For this to become clearer, let us
set $d\mathrm{U} = \mathrm{K} dx + k dy $; thus
$\frac{d\mathrm{U} }{dx} = \mathrm{K} $ and $\frac{d\mathrm{U} }{dy} =
k$, so that $\frac{d\mathrm{U} }{dx} . dx + \frac{d\mathrm{U} }{dy} =
\mathrm{K} dx + k dy = d\mathrm{U} $. Since its integral is $
\mathrm{U} = \frac{d\mathrm{S} }{dt} $, we shall have
\begin{equation*}
dp = \alpha dx - 2 u du - 2 v dv - 2 d\mathrm{U} 
\end{equation*}
from where it appears by integration:
\begin{equation*}
p = \mathrm{Const.} + \alpha x - u u - v v - \frac{2 d\mathrm{S} }{dt} 
\end{equation*}
with a given function S of the coordinates $x$, $y$ and $t$ themselves, whose differential, for $dt=0$ is $u dx + v dy$. 

{\bf 52}.~~~In order to understand better the nature of these formulas, let us consider the true velocity of the point $l$, which is $\mathrm{V} = \sqrt( u u + v v )$. And the pressure will be: 
$p = \mathrm{Const.} + \alpha x - \mathrm{V V} - \frac{2d\mathrm{S} }{dt}$ : in which the last term 
$d\mathrm{S} $ denotes the differential of $\mathrm{S} = \int (u dx + v dy ) $
itself, where the time $t$ is allowed to vary.

{\bf 53}.~~~If we now wish to also take into account friction, let us
set it proportional to the pressure $p$. While the point $l$ travels
the element $ds$,  the retarding force arising from the friction is
$=\frac{p}{f} $; so that, setting $\frac{d\mathrm{S}}{dt} = \mathrm{U}
$, our differential equation will be for constant $t$
\begin{equation*}
dp = \alpha dx - \frac{p}{f} ds - \mathrm{V} d\mathrm{V} - 2 d\mathrm{U} ,
\end{equation*}
from where we obtain by integration, taking $e$ for the number whose
hyperbolic\footnote{Natural} logarithm is $=1$,
\begin{equation*}
\begin{split}
& p = e^{\frac{-s}{f} } \int e^{\frac{s}{f} } (\alpha dx - 2 \mathrm{V} d\mathrm{V} - 2 d\mathrm{V} ) \quad \mathrm{or} \\
& p = \alpha x - \mathrm{V V} - 2 \mathrm{U} - \frac{1}{f} e^{\frac{-s}{f} } \int e^{\frac{s}{f} } (\alpha x - 
\mathrm{V V} - 2 \mathrm{U} ) ds .
\end{split}
\end{equation*}
 
{\bf 54}.~~~The criterion of the motion which drives the fluid in
 reality consists in this that, fixing the time $t$, the differential
 $u dx + v dy $ has to be complete: also continuity and constant
 uniform density demand that $\frac{du}{dx} + \frac{dv}{dy} = 0 $,
 hence it follows too that this differential $u dy - v dx $ will have
 to be complete.\footnote{The published version has $u dx + v dy $,
 a mistake not present in Euler, 1752.} From where
both velocities $u$ and $v$ jointly must be functions of the
 coordinates $x$ and $y$ with the time $t$ in such a way that both
 differential formulas $u dx + v dy $ and $u dy - v dx
 $\footnote{Previous mistake repeated in the published version.} be
 complete differentials.
 
{\bf 55}.~~~Let us set up the same investigation in general, giving
the point $\lambda$ three velocities directed parallel to the axes AL,
AB, AC. Let $u$, $v$, $w$ denote these functions, which depend on
coordinates $x$, $y$, $z$, besides $t$. After a differentiation
we obtain
\begin{equation*}
\begin{split}
& d u = \mathrm{L} dx + l dy + \lambda dz + \mathfrak{L} dt \\
& d v = \mathrm{M} dx + m dy + \mu dz + \mathfrak{M} dt \\
& d w = \mathrm{N} dx + n dy + \nu dz + \mathfrak{N} dt. 
\end{split}
\end{equation*}
Although here the time $t$ is also taken as variable, nonetheless
for the motion to be possible, by the preceding
condition\footnote{From Part I.} we have
$\mathrm{L} + m + \nu = 0$, or, which reexpresses the same
\begin{equation*}
\frac{du}{dx} + \frac{dv}{dy} + \frac{dw}{dz} = 0, 
\end{equation*}
a condition on which the present examination  does not depend. 

{\bf 56}.~~~After the passage of time interval $dt$ the point $\lambda
$ is carried to $\pi $, and it travels a distance $u dt $ parallel to
the axis AL, a distance $v dt $ parallel to the axis AB and a
distance $w dt $ parallel to the axis AC. Thus the three
velocities of the point which has moved from $\lambda $ to $\pi $ will be:
\begin{equation*}
\begin{split}
&  
\mathrm{parallel\ to} \ \makebox[3.2ex][l]{$\mathrm{AL}$} 
= \makebox[1.6ex][l]{$u$} + \makebox[1.8ex][l]{$\mathrm{L}$}\makebox[1.6ex][l]{$u$}
\, dt + \makebox[1.6ex][l]{$l$}\makebox[1.6ex][l]{$v$} \, dt + 
\makebox[1.6ex][l]{$\lambda $}\makebox[1.6ex][l]{$w$} \,  dt + 
\makebox[1.8ex][l]{$\mathfrak{L}$} \, dt  ; \\ 
&  
\mathrm{parallel\ to} \ \makebox[3.2ex][l]{$\mathrm{AB}$} 
= \makebox[1.6ex][l]{$v$} + \makebox[1.8ex][l]{$\mathrm{M}$}\makebox[1.6ex][l]{$u$} 
\, dt + \makebox[1.6ex][l]{$m$}\makebox[1.6ex][l]{$v$} \, dt + 
\makebox[1.6ex][l]{$\mu $}\makebox[1.6ex][l]{$w$} \, dt + 
\makebox[2.2ex][l]{$\mathfrak{M}$} \, dt  ;\\ 
&  
\mathrm{parallel\ to} \ \mathrm{AC} 
= \makebox[1.6ex][l]{$w$} + \makebox[1.8ex][l]{$\mathrm{N}$} \makebox[1.6ex][l]{$u$} 
\,dt + \makebox[1.6ex][l]{$n$}\makebox[1.6ex][l]{$v$} \, dt + 
\makebox[1.6ex][l]{$\nu $}\makebox[1.6ex][l]{$w$} \, dt + 
\makebox[2.0ex][l]{$\mathfrak{N}$} \, dt  ,
\end{split}
\end{equation*}
and the accelerations parallel to the same directions will be 
\begin{equation*}
\begin{split}
& 
\mathrm{par.} \ \makebox[3.2ex][l]{$\mathrm{AL}$} 
= 2 ( \makebox[1.6ex][l]{$\mathrm{L}$}\makebox[1.6ex][l]{$u$} + 
\makebox[1.6ex][l]{$l$}\makebox[1.6ex][l]{$v$} + \makebox[1.6ex][l]{$\lambda $}
\makebox[1.6ex][l]{$w$} + \makebox[1.8ex][l]{$\mathfrak{L}$} )  ; \\
& 
\mathrm{par.} \ 
\makebox[3.2ex][l]{$\mathrm{AB}$} 
= 2 ( \makebox[1.6ex][l]{$\mathrm{M}$}\makebox[1.6ex][l]{$u$} + 
\makebox[1.6ex][l]{$m$}\makebox[1.6ex][l]{$v$} + \makebox[1.6ex][l]{$\mu $}
\makebox[1.6ex][l]{$w$} + \makebox[2.2ex][l]{$\mathfrak{M}$} ) ;  \\
& 
\mathrm{par.} \ \makebox[3.2ex][l]{$\mathrm{AC}$} = 2 ( 
\makebox[1.6ex][l]{$\mathrm{N}$}\makebox[1.6ex][l]{$u$} + \makebox[1.6ex][l]{$n$} 
\makebox[1.6ex][l]{$v$} + \makebox[1.6ex][l]{$\nu$}\makebox[1.6ex][l]{$w$} + 
\makebox[2.0ex][l]{$\mathfrak{N}$} ) .
\end{split}
\end{equation*}

{\bf 57}.~~~If we take the axis AC to be vertical, in such a way that
the remaining two AL and AB are horizontal, the accelerating force due
to gravity arises parallel to the axis $\mathrm{AC}$ with the value $-1$.  Then
indeed, denoting the pressure at $\lambda$ by $ p$, its differential,
at constant time is
\begin{equation*}
d p = \mathrm{R} \, dx + r dy + \rho dz ,
\end{equation*}
from which we obtain  the three accelerating forces
\begin{equation*}
\mathrm{par.} \ \mathrm{AL} = \mathrm{R} ; \quad \mathrm{par.} \ \mathrm{AB} = -r ; 
\quad \mathrm{par.} \ \mathrm{AC} = - \rho 
\end{equation*}
which are in fact easily collected in the same manner as was done in
$\S\S.~44$ and 45, so that it is not
necessary to repeat the same computation. Hence we obtain the
following equations\footnote{These are the three dimensional Euler
equations.}
\begin{equation*}
\begin{split}
& \mathrm{R} = - 2 (\mathrm{L} u + l v + \lambda w  + \mathfrak{L} )  \\
& r = - 2 (\mathrm{M} u + m v + \mu w + \mathfrak{M} )  \\
& \rho = -1 - 2 (\mathrm{N} u + n v + \nu w + \mathfrak{N} ) 
\end{split}
\end{equation*}

{\bf 58}.~~~Since the differential formula $dp = \mathrm{R} dx + r dy + \rho dz $
has to be a complete differential, we have
\begin{equation*}
\frac{d\mathrm{R} }{dy} = \frac{dr}{dx} ; \quad \frac{d\mathrm{R} }{dz} = \frac{d\rho }{dx} ; \quad 
\frac{dr}{dz} = \frac{d\rho }{dy } .
\end{equation*} 
After a differentiation and a division by
$-2$ the following three equations are
obtained\footnote{The printed version contains  mistakes not present
  in Euler, 1752: in the formula labelled II, instead of $L$
there is  $\mathfrak{L} $;  in the formula labelled III there is a $v$ instead
of $u$.}
{\small
\begin{equation*}
\begin{split}
&  \makebox[3.2ex][l]{I}   \! \! 
 \left\{  \begin{array}{ l l l l l l l l l l l l l l }
 \frac{u d\mathrm{L} }{dy} & + & \frac{v dl}{dy}    & + & 
 \frac{w d \lambda }{dy} & + & \frac{d\mathfrak{L} }{dy} 
 & + & \mathrm{L} l  & +  & lm & +  & \lambda n  & = \\
 \frac{u d\mathrm{M} }{dx} & + & \frac{v dm}{dx} & + & \frac{w d\mu }{dx}         
 & +  & \frac{d\mathfrak{M} }{dx} 
 & + & \mathrm{ML} & + & m\mathrm{M} & + & \mu \mathrm{N}        &    
 \\ \end{array}    \right. \\
 &  \makebox[3.2ex][l]{II}  \! \!  
 \left\{  \begin{array}{ l l l l l l l l l l l l l l }
 \frac{u d\mathrm{L} }{dz}  & + & \frac{v dl}{dz}   & + & 
 \frac{w d \lambda }{dz} & + & \frac{d\mathfrak{L} }{dz} 
 & + & \mathrm{L} \lambda  & +  & l \mu & +  & \lambda \nu & = \\
 \frac{u d\mathrm{N} }{dx} & + & \frac{v dn}{dx} & + & \frac{w d\nu }{dx}            
 & +  & \frac{d\mathfrak{N} }{dx} 
 & + & \mathrm{NL} & + & n\mathrm{M} & + & \nu \mathrm{N}       &    \\ 
 \end{array}    \right.   \\
 &  \makebox[3.2ex][l]{III}  \! \! 
\left\{  \begin{array}{ l l l l l l l l l l l l l l }
 \frac{u d\mathrm{M} }{dz} & + & \frac{v dm}{dz}   & + & \frac{w d \mu }{dz} & + 
 & \frac{d\mathfrak{M} }{dz} 
 & + & \mathrm{M} \lambda & +  & m \mu & +  & \mu \nu & = \\
 \frac{u d\mathrm{N} }{dy} & + & \frac{v dn}{dy} & + & \frac{w d\nu }{dy}         
 & +  & \frac{d\mathfrak{N} }{dy} 
 & + & \mathrm{N}l & + & n m & + & \nu n       &    \\ \end{array}    \right. 
\end{split}
\end{equation*}
}

{\bf 59}.~~~Moreover, because of the nature of the complete differentials, we 
have
\begin{equation*}
\begin{split}
& 
\frac{d\makebox[1.6ex][l]{L} }{dy} = \frac{d\makebox[1.6ex][l]{$l$} }{dx} ; 
\quad   
\frac{d\makebox[1.6ex][l]{$m$} }{dx} = \frac{d\makebox[1.6ex][l]{M} }{dy} ; 
\quad
\frac{d\makebox[1.6ex][l]{$\lambda $} }{dy} = 
\frac{d\makebox[1.6ex][l]{$l$} }{dz} ; \\   
& 
\frac{d\makebox[1.6ex][l]{$\mu $ } }{dx} = 
\frac{d\makebox[1.6ex][l]{M} }{dz} ; 
\quad 
\frac{d\makebox[1.6ex][l]{$\mathfrak{L} $ } }{dy} = 
\frac{d\makebox[1.6ex][l]{$l$} }{dt} ; 
\quad   
\frac{d\makebox[1.6ex][l]{$\mathfrak{M} $} }{dx} = 
\frac{d\makebox[1.6ex][l]{M} }{dt}  
\\
& 
\frac{d\makebox[1.6ex][l]{L} }{dz} = 
\frac{d\makebox[1.6ex][l]{$\lambda $ } }{dx} ; 
\quad   
\frac{d\makebox[1.6ex][l]{$l$} }{dz} = 
\frac{d\makebox[1.6ex][l]{$\lambda $} }{dy} ; 
\quad 
\frac{d\makebox[1.6ex][l]{$n$} }{dx} = 
\frac{d\makebox[1.6ex][l]{N} }{dy} ; \\
& 
\frac{d\makebox[1.6ex][l]{$\nu $} }{dx} = 
\frac{d\makebox[1.6ex][l]{N} }{dz} ; 
\quad 
\frac{d\makebox[1.6ex][l]{$\mathfrak{L}$} }{dz} = 
\frac{d\makebox[1.6ex][l]{$\lambda $} }{dt} ; 
\quad   
\frac{d\makebox[1.6ex][l]{$\mathfrak{N}$} }{dx} = 
\frac{d\makebox[1.6ex][l]{N} }{dt}  ; \\
& 
\frac{d\makebox[1.6ex][l]{M} }{dz} = 
\frac{d\makebox[1.6ex][l]{$\mu $} }{dx} ; 
\quad   
\frac{d\makebox[1.6ex][l]{N} }{dy} = 
\frac{d\makebox[1.6ex][l]{$n$} }{dx} ; 
\quad 
\frac{d\makebox[1.6ex][l]{$m$} }{dz} = 
\frac{d\makebox[1.6ex][l]{$\mu $} }{dy} ; \\
& 
\frac{d\makebox[1.6ex][l]{$\nu $} }{dy} = 
\frac{d\makebox[1.6ex][l]{$n$} }{dz} ; 
\quad 
\frac{d\makebox[1.6ex][l]{$\mathfrak{M} $} }{dz} = 
\frac{d\makebox[1.6ex][l]{$\mu $} }{dt} ; 
\quad   
\frac{d\makebox[1.6ex][l]{$\mathfrak{N} $} }{dy} = 
\frac{d\makebox[1.6ex][l]{$n$} }{dt} ,
\end{split}
\end{equation*}
after substituting of which values those three equations will be
transformed into these\footnote{These are the equations for the
vorticity.}
\begin{equation*}
\begin{split}
& \left( \frac{d l - d \mathrm{M} }{dt} \right) +  u  \left( \frac{d l - d \mathrm{M} }{dx} \right) + 
v  \left( \frac{d l - d \mathrm{M} }{dy} \right) + \\
& w  \left( \frac{d l - d \mathrm{M} }{dz} \right) + 
(l - \mathrm{M} ) (\mathrm{L} + m ) + \lambda n - \mu \mathrm{N} = 0 , \\
& \left( \frac{d \lambda - d \mathrm{N} }{dt} \right) +  u  \left( \frac{d \lambda - d \mathrm{N} }{dx} \right) + 
v  \left( \frac{d \lambda - d \mathrm{N} }{dy} \right) + \\
& w  \left( \frac{d \lambda - d \mathrm{N} }{dz} \right) + 
(\lambda - \mathrm{N} ) (\mathrm{L} + \nu ) + l \mu  - n \mathrm{M} = 0 , \\
& \left( \frac{d \mu - d n }{dt} \right) +  u  \left( \frac{d \mu - d n }{dx} \right) + 
v  \left( \frac{d \mu - d n }{dy} \right) + \\
& w  \left( \frac{d \mu - d n }{dz} \right) + 
(\mu - n ) (m + \nu ) + \mathrm{M} \lambda  - \mathrm{N} l = 0 .
\end{split}
\end{equation*}

{\bf 60}.~~~Now it is manifest that these three equations are
satisfied by the following three values
\begin{equation*}
l = \mathrm{M} ; \quad \lambda = \mathrm{N} ; \quad \mu = n 
\end{equation*}
in which is contained the criterion furnished by the consideration of
the forces. Here therefore follows that in the notation chosen we
have\footnote{Here Euler repeats the mistake of assuming that the only
  solution is zero-vorticity flow; in Euler, 1755 this will be corrected.}
\begin{equation*}
\frac{du}{dy} = \frac{dv}{dx} ; \quad \frac{du}{dz} = \frac{dw}{dx} ; \quad 
\frac{dv}{dz} = \frac{dw}{dy} 
\end{equation*}
these conditions moreover are the same as those which are required in
order that the formula $u dx + v dy + w dz $ be a complete
differential. From which this criterion consists in that the three
velocities $u$, $v$ and $w$ have to be functions of $x$, $y$ and $z$
together with $t$ in such a manner that for fixed constant time the
formula $u dx + v dy + w dz $ admits an integration.

{\bf 61}.~~~Taking the time $t$ constant or $dt  = 0 $, we have
\begin{equation*}
\begin{split}
& d u = \mathrm{L} dx + \mathrm{M} dy + \mathrm{N} dz \\ 
& d v = \mathrm{M} dx + m dy + n dz  \\
& d w = \mathrm{N} dx + n dy + \nu dz 
\end{split}
\end{equation*}
moreover, for R, $r$ and $\rho $ the values are
\begin{equation*}
\begin{split}
& \mathrm{R} = - 2 ( \mathrm{L} u + \mathrm{M} v + \mathrm{N} w + \mathfrak{L} )  \\
& r =  - 2 ( \mathrm{M} u + m v + n w + \mathfrak{M} )  \\ 
& \rho = - 1 - 2 (\mathrm{N} u + n v + \nu w + \mathfrak{N} ) 
\end{split}
\end{equation*}
Regarding the pressure $p$, we obtain the following equation
\begin{equation*}
\begin{split}
& dp = - d z \\
& - 2 u  (\mathrm{L} dx + \mathrm{M} dy + \mathrm{N} dz ) = - dz - 2 u d u - 2 v d v - 2 w d w    \\
& - 2 v ( \mathrm{M} d x + m d y + n dz ) \quad \qquad \  \  
- 2 \mathfrak{L}  dx - 2 \mathfrak{M} - 2 
\mathfrak{N} dz    \\
& - 2 w ( \mathrm{N} dx + n dy + \nu dz ) \\
 & - 2 \mathfrak{L} dx - 2 \mathfrak{M} d y - 2 \mathfrak{N} dz 
\end{split}
\end{equation*}

{\bf 62}.~~~Since in truth $\mathfrak{L} = \frac{du}{dt} $; $\mathfrak{M} = \frac{dv}{dt} $; 
$\mathfrak{N} = \frac{dw}{dt} $, we obtain by integration
\begin{equation*}
p = \mathrm{C} - z - uu - vv - ww - 2 \int \left( \frac{du}{dt} dx + \frac{dv}{dt} dy + \frac{dw}{dt} dz \right)
\end{equation*}
By the previously ascertained condition  $u d x + v d y + w d z $
is integrable. Let us denote its integral by
$ \mathrm{S} $, which can also involve the time $t$; taking also the
time $t$ variable, we have
\begin{equation*}
d \mathrm{S} = u dx + v d y + w d z + \mathrm{U} dt ,
\end{equation*}
and we have $\frac{ d u}{ dt } = \frac{d \mathrm{U} }{ dx } $;
 $\frac{dv}{dt} = \frac{d \mathrm{U} }{dy} $; $ \frac{dw}{dt} =
 \frac{d \mathrm{U} }{dz} $. From where, with time generally taken
 constant, it can be
 assumed in the above integral that 
 \begin{equation*}
 \frac{d\mathrm{U} }{dx } \, dx + \frac{d\mathrm{U} }{dy} \, dy + 
 \frac{d\mathrm{U} }{dz} \, dz = d\mathrm{U} ,
 \end{equation*} 
and we obtain\footnote{The published version has a $ds$ in the
denominator, instead of the correct $dt$, found in Euler, 1752.}
\begin{equation*}
 \begin{split}
 & p = \mathrm{C}  - z - uu - vv - ww - 2 \mathrm{U} , \quad \mathrm{or} \\
 & p = \mathrm{C}  - z - uu - vv - ww - 2 \frac{d\mathrm{S} }{dt} 
 \end{split}
\end{equation*}

{\bf 63}.~~~Thus, $uu + vv + ww$ is manifestly expressing the square
of the true velocity of the point $\lambda $, so that, if the true
velocity of this point is denoted  $V$, the following equation is
obtained for the pressure\footnote{This is basically the Bernoulli pressure 
law for potential flow.}
\begin{equation*}
p = \mathrm{C} - z - \mathrm{VV} - \frac{2 d\mathrm{\mathrm{S} } }{dt}.
\end{equation*}
To use this, firstly one must seek the integral S of the formula $u
dx + v dy + w dz $ which should be complete. This is differentiated
again, taking only the time $t$ as variable. After division by $dt$,
one obtains the value of the formula $\frac{d\mathrm{S}
}{dt} $, which enters into the expression for the state of the
pressure $p$.

{\bf 64}.~~~But before we may add here the previous criterion,
regarding possible motion,  the three
velocities $u$, $v$ and $w$ must be such functions of the three
coordinates $x$, $y$ and $z$, and of time $t$ that,  firstly,  $u dx
+ v dy + w dz $ be a complete differential and, secondly,  that
the condition $\frac{du}{dx} + \frac{dv}{dy} + \frac{dw}{dz} = 0$
holds.  The whole motion of fluids endowed with
invariable density is subjected to these two conditions.

Furthermore,  if we  take also the time $t$ to be variable, and the 
differential formula $u dx + v dy + w dz + \mathrm{U} dt $
is a complete differential,  the state of the pressure at any point 
$\lambda $, expressed as an altitude $p$,  will be given by
 \begin{equation*}
 p = \mathrm{C} - z - uu - vv - ww - 2 \mathrm{U} ,
 \end{equation*}
 if only the fluid enjoys the natural gravity and the plane BAL is horizontal. 

{\bf 65}.~~~Suppose we had attributed another direction to the gravity or
even adopted arbitrary variable forces acting on the particles of the
fluid. Differences would arise in the values of the pressure, but the law
which the three velocities of the fluid have to obey would not suffer any
changes. Thus, whatever the acting forces, the three velocities $u$, $v$ and
$w$ have to satisfy the conditions that the differential formula $ u dx + v dy
+ w dz$ be complete and that $\frac{du}{dx} + \frac{dv}{dy} + \frac{dw}{dz} =
0 $ should hold.  Therefore, the three velocities $u$, $v$ and $w$ can be
fixed in infinitely many ways while satisfing the two conditions; and then it
is possible to prescribe the pressure at every point of the
fluid.\footnote{Many statements in this paragraph are rendered invalid by the
generally incorrect assumption of potential flow.}

{\bf 66}.~~~However, much more difficult would be the following question:
given the acting forces and the pressure at all places, to determine the
motion of the fluid at all points.  Indeed, we would then have some
equations\footnote{The plural is here used probably because this relation has
to be satisfied at all points.} of the form $p = \mathrm{C} - z - uu - vv - ww
- 2 \mathrm{U} $, from which the relation of the functions $u$, $v$ and $w$
would have to be defined in such a way that not only the equations themselves
would be satisfied , but also the previously contributed rules\footnote{Incompressibility and potentiality.} 
would have to be obeyed; this work would certainly require the greatest force
of calculation. It is fitting therefore to inquire in general into the nature
of functions proper to satisfy both criteria.
 
{\bf 67}.~~~Most conveniently therefore let us begin with the
characterization
of the 
integral quantity S, whose differential is $u dx + v dy + w dz $, when
time is held constant. Let thus S
be a function of $x$, $y$ and $z$, the time $t$ being contained
in constant quantities.  When S is differentiated,
the coefficients of the differentials $dx$, $dy$ and $dz$ are
the velocities $u$, $v$ and $w$ which at the present
time suit the point of fluid $\lambda $, whose coordinates
are $x$, $y$ and $z$.  The question thus arises here to find
the  functions S of $x$, $y$ and $z$ such that
$\frac{du}{dx} + \frac{dv}{dy} +
\frac{dw}{dz} = 0 $; now, since we have $ u = \frac{dS}{dx} $, $v =
\frac{dS}{dy} $ and $w = \frac{dS}{dz} $  it follows that $\frac{dd \mathrm{S}
}{d x^2 } + \frac{dd \mathrm{S} }{dy^2 } + \frac{dd \mathrm{S} }{dz^2
} = 0 $.\footnote{This is what will later be called Laplace's equation.}

{\bf 68}.~~~Since it is not plain how this can be handled in general,
I shall 
consider certain rather general cases. Let
\begin{equation*}
\mathrm{S} = (\mathrm{A} x + \mathrm{B} y + \mathrm{C} z )^n.
\end{equation*}
We  have
\begin{equation*}
\begin{split}
&  \frac{d \mathrm{S} }{dx} = n \mathrm{A} ( \mathrm{A} x + \mathrm{B} y + \mathrm{C} z )^{n-1} \ \mathrm{and} \\ 
&  \frac{dd \mathrm{S} }{d x^2 } = n (n-1) \mathrm{AA} (\mathrm{A} x + \mathrm{B} y + \mathrm{C} z)^{n-2} 
\end{split}
\end{equation*}
and the expressions for $\frac{dd \mathrm{S} }{dy^2 } $ and $\frac{dd
  \mathrm{S} }{dz^2 } $ will be similar.  Thus we have to satisfy
\begin{equation*}
n (n - 1) ( \mathrm{A} x + \mathrm{B} y + \mathrm{C} z)^{n-2} 
( \mathrm{AA} + \mathrm{BB} + \mathrm{CC}) = 0 
\end{equation*}
which is plainly satisfied when either $n=0$ or $n=1$. Thus we have
the solutions $\mathrm{S} = \mathrm{Const.}$ and $\mathrm{S} =
\mathrm{A} x +  \mathrm{B} y + \mathrm{C} z $, where the constants $A$,
$B$ and $C$ are arbitrary.

{\bf 69}.~~~But if $n$ is neither $0$, nor $1$, we necessarily have: 
$\mathrm{AA} + \mathrm{BB} + \mathrm{CC} = 0$: and then S is given by
\begin{equation*}
\mathrm{S} = (\mathrm{A} x + \mathrm{B} y + \mathrm{C} z )^{n}
\end{equation*}
for any value of the exponent $n$;  even the time $t$ itself will
possibly enter in $n$. Furthermore we can add up  arbitrarily  many
such  S 
and obtain yet another solution.\footnote{In modern terms, Euler is
  here using the linear character of the Laplace equation.}  The function 
\begin{equation*}
\begin{split}
& \mathrm{S} = \alpha + \beta x + \gamma y + \delta z + \epsilon ( \mathrm{A} x + \mathrm{B} y + 
\mathrm{C} z )^n +
\\
& \zeta ( \mathrm{A} ^{\prime } x + \mathrm{B} ^{\prime } y + \mathrm{C} ^{\prime } z )^{n^{\prime } } +  
\eta   ( \mathrm{A} ^{\prime \prime } x + \mathrm{B} ^{\prime \prime } y + \mathrm{C} ^{\prime \prime } z )^{n^{\prime \prime } }  + \\
& \theta  ( \mathrm{A} ^{\prime \prime \prime } x + 
\mathrm{B} ^{\prime \prime \prime } y + \mathrm{C} ^{\prime \prime \prime } z )^{n^{\prime \prime \prime } } 
\ \mathrm{etc.}
\end{split}
\end{equation*}
will satisfy the condition only if we have:
\begin{equation*}
\begin{split}
& \mathrm{A} \mathrm{A} + \mathrm{B} \mathrm{B} + \mathrm{C} \mathrm{C} = 0 ; \quad 
\mathrm{A} ^{\prime } \mathrm{A} ^{\prime } + \mathrm{B} ^{\prime } \mathrm{B} ^{\prime } + 
\mathrm{C} ^{\prime } \mathrm{C} ^{\prime } = 0 ; \\
&  \mathrm{A} ^{\prime \prime } 
\mathrm{A} ^{\prime \prime } + \mathrm{B} ^{\prime \prime } \mathrm{B} ^{\prime \prime } + 
\mathrm{C} ^{\prime \prime } \mathrm{C} ^{\prime \prime } = 0 \ \mathrm{etc.}
\end{split}
\end{equation*}

{\bf 70}.~~~Here suitable values are given for $S$ in which the coordinates
$x$, $y$, $z$  have either one, or two, or three, or four
dimensions\footnote{In modern terms we would say ``which are
  polynomials in $x$, $y$, $z$ of degrees up to four''.}
\begin{equation*}
\begin{split}
\mathrm{I.}  \ 
& \mathrm{S} = \mathrm{A} \\
\mathrm{II.} \ 
& \mathrm{S} = \mathrm{A} x + \mathrm{B} y + \mathrm{C} z  \\
\mathrm{III.} \ 
& \mathrm{S} = \mathrm{A} xx + \mathrm{B} yy + \mathrm{C} zz + 2 \mathrm{D} xy + 2 \mathrm{E} x z +
2 \mathrm{F} y z \\
\mathrm{with} 
& \quad \mathrm{A} + \mathrm{B} + \mathrm{C} = 0 \\
\mathrm{IV.} \ 
& \mathrm{S} = \mathrm{A} x^3 + \mathrm{B} y^3 + \mathrm{C} z^3  + 3 \mathrm{D} 
x x y + 3 \mathrm{F} x x z + \mathrm{H} y y z + \\
& 6 \mathrm{K} x y z + 3 \mathrm{E} x y y + 3 \mathrm{G} x z z + 3 \mathrm{I} y z z \\
& \mathrm{with} \quad \mathrm{A} + \mathrm{E} + \mathrm{G} = 0 ; \quad 
\mathrm{B} + \mathrm{D} + \mathrm{I} = 0 ; \\ 
& \mathrm{C} + \mathrm{F} + \mathrm{H} = 0 \\
\mathrm{V.}   \\
                        & + \mathrm{A} x^4 + 6 \mathrm{D} xx yy + 4 \mathrm{G} x^3 y + 4 \mathrm{H} 
                         x y^3 + 12 \mathrm{N} x x y z \\
\mathrm{S} = & + \mathrm{B} y^4 + 6 \mathrm{E} x x z z + 4 \mathrm{I} x^3 z + 4 \mathrm{K} x z^3 + 
                            12 \mathrm{O} x y y z  \\
                         & + \mathrm{C} z^4 + 6 \mathrm{F} y y z z + 4 \mathrm{L} y^3 z + 4 \mathrm{M} y z^3 + 
                             12 \mathrm{P} x y z z \\
\mathrm{with}  \quad
& \mathrm{A} + \mathrm{D} + \mathrm{E} = 0   \quad \mathrm{G} + \mathrm{H} + \mathrm{P} = 0 \\                                                    
& \mathrm{B} + \mathrm{D} + \mathrm{F} = 0   \quad \mathrm{I} + \mathrm{K} + \mathrm{O} = 0 \\
& \mathrm{C} + \mathrm{E} + \mathrm{F} = 0   \quad \mathrm{L} + \mathrm{M} + \mathrm{N} = 0 
\end{split}
\end{equation*}

{\bf 71}.~~~Hence it is clear how these formulas are to be gotten for any
order. First, simply give to the various terms the numerical coefficients
which belong to them from the law of permutation, or, equivalently, which
arise when the trinomial $x + y + z$ is raised to that same power. Let
indefinite letters $A$, $B$, $C$, etc., be adjoined to the numerical
coefficients. Then, ignoring the coefficients, observe whenever there occur
three terms of the type $\mathrm{L} \mathrm{Z} x ^2 + \mathrm{M} \mathrm{Z}
y^2 + \mathrm{N} \mathrm{Z} z^2 $ having a common factor $\mathrm{Z} $ formed
from the variables. Whenever this occurs, set the sum of the literal
coefficients $\mathrm{L} + \mathrm{M} + \mathrm{N} $ equal to zero. For 
example, for the fifth power we have 
\begin{equation*}
\begin{split}
\mathrm{S} =  & \mathrm{A} x^5 + 5 \mathrm{D} x^4 y +  5 \mathfrak{D} x^4 z 
                    + 10 \mathrm{G} x^3 y y + \mathfrak{G} x^3 z z + \\
              & 20 \mathrm{K} x ^3 y z + 30 \mathrm{N} x y y z z + \\
              & \mathrm{B} x^5 + 5 \mathrm{E} x^4 y +  5 \mathfrak{E} x^4 z  
                + 10 \mathrm{H} x^3 y y + \mathfrak{H} x^3 z z + \\
              & 20 \mathrm{L} x ^3 y z + 30 \mathrm{O} x y y z z \\
              & + \mathrm{C} x^5 + 5 \mathrm{F} x^4 y +  5 \mathfrak{F} x^4 z  
                + 10 \mathrm{I} x^3 y y + \mathfrak{I} x^3 z z + \\
              & 20 \mathrm{M} x ^3 y z + 30 \mathrm{P} x y y z z \\
\end{split}
\end{equation*}
and the following determinations of the coefficient letters are obtained
\begin{equation*}
\begin{split}
& \mathrm{A} + \mathrm{G} + \mathfrak{G} = 0 ;  \quad \mathrm{D} + \mathrm{H} + \mathrm{O} = 0 ;
\quad \mathfrak{D} + \mathrm{I} + \mathrm{P} = 0  ; \\ 
& \mathrm{B} + \mathrm{H} + \mathfrak{H} = 0 ;  \quad \mathrm{E} + \mathrm{G} + \mathrm{N} = 0 ;
\quad \mathfrak{G} + \mathrm{F} + \mathrm{P} = 0 ;  \\
& \mathrm{K} + \mathrm{L} + \mathrm{M}  = 0 ; \\
& \mathrm{C} + \mathrm{I} + \mathfrak{F} = 0  ; \quad \mathrm{F} + \mathfrak{G} + \mathrm{N} = 0 ;
\quad \mathfrak{F} + \mathfrak{H} + \mathrm{O} = 0 .
\end{split}
\end{equation*}
In the same way for the sixth order such determinations will give $15$, for the seventh $21$, for the eighth $28$ and so on. 
 
{\bf 72}.~~~In the very first formula $\mathrm{S} = \mathrm{A} $ the coordinates
$x$, $y$ and $z$ are clearly not intertwined.  Thus the three
velocities $u$, $v$ and $w$ are equal to zero, and hence this  describes a quiet
state of fluid. Also the pressure at an arbitrary point for different times
will be able to vary in an arbitrary manner. Indeed $A$ is an arbitrary
function of time and,  for a given time $t$, the pressure at the point 
$\lambda $ is $p = \mathrm{C} - \frac{2 d\mathrm{A} }{dt} - z $. Through
this formula is revealed the state of the fluid, when it is subjected
at an arbitrary instant to arbitrary forces, which nevertheless balance
each other, so that no motion in the fluid can arise from them: where it
happens, if the fluid is enclosed in a vase from which it can nowhere
escape, it is also compressed by suitable forces inside.

{\bf 73}.~~~Moreover, the second formula $\mathrm{S} = \mathrm{A} x + \mathrm{B} y +
    \mathrm{C} z $, after differentiation, gives these three velocities to
    the point $\lambda $:
\begin{equation*}
u = \mathrm{A} ; \quad v = \mathrm{B} \quad \mathrm{and} \quad w = \mathrm{C} .  
\end{equation*}
Thus simultaneously, all points of the fluid are carried by an identical
motion in the same direction.  From which the whole fluid moves in the same
manner as a solid body, carried only by a forward motion. But at different
times the velocities as well as the direction of this motion are able to be
varied in an arbitrary way, depending on what the extrinsic driving forces
require. Therefore, the pressure at the point $\lambda $ at the time $t$ on
which $\mathrm{A} $, $\mathrm{B} $, $\mathrm{C} $ depend, is\footnote{The
printed version, but not Euler, 1752, has a missing $\mathrm{B} \mathrm{B} $
in the formula.} $p = \mathrm{C} - z - \mathrm{A} \mathrm{A} - \mathrm{B}
\mathrm{B} - \mathrm{C} \mathrm{C} - 2 x \frac{d\mathrm{A} }{dt} - 2 y \frac{d
\mathrm{B} }{dt} - 2 z \frac{d \mathrm{C} }{dt} $.

{\bf 74}.~~~The third formula $\mathrm{S} = \mathrm{A} x x + \mathrm{B} y y +
\mathrm{C} z z + 2 \mathrm{D} x y + 2 \mathrm{E} x z + 2 \mathrm{F} y z$,
where $\mathrm{A} + \mathrm{B} + \mathrm{C} = 0 $, gives the following three
velocities\footnote{In both the printed version and in Euler, 1752, the first
  velocity component is mistakenly denoted by $\alpha $.}  of the point
$\lambda $: $u = 2 \mathrm{A} x + 2 \mathrm{D} y + 2 \mathrm{E} z $; $v= 2
\mathrm{B} y + 2 \mathrm{D} x + 2 \mathrm{F} z $; $w = 2 \mathrm{C} z + 2
\mathrm{E} x + 2 \mathrm{F} y $, or $w = 2 \mathrm{E} x + 2 \mathrm{F} y - 2
(\mathrm{A} + \mathrm{B} ) z $. Here, at a given instant, different points of
the fluid are carried by different motions; moreover, in the time development
an arbitrary motion of a given point is permitted, because $\mathrm{A} $,
$\mathrm{B} $, $\mathrm{D} $, $\mathrm{E} $, $\mathrm{F} $ can be arbitrary
functions of the time $t$. Finally, a much greater variety can take place, if
more elaborate values are given to the function $\mathrm{S} $.

{\bf 75}.~~~In the second case the motion of the fluid was corresponding to the
    forward motion of a solid body in which, plainly, at any instant the
    different parts are carried by a motion equal and parallel to itself.  In
    other cases the motion of the fluid could be suspected to correspond to
    solid-body motion, either rotational or anomalous. 
It suffices to
put forward such a hypothesis -- beyond the second case -- to find
that it cannot take place. Indeed, 
    in order to happen, not only would it be necessary that the pyramid $\pi
    \Phi \rho \sigma $ would be equal,\footnote{In volume.} but also similar to the
    pyramid $\lambda \mu \nu o $, or that the following  holds
\begin{equation*}
\begin{split}
& \pi \Phi = \lambda \mu = dx = \sqrt{} ( \mathrm{Q} \mathrm{Q} + q q + \Phi \Phi ) \\ 
& \pi \rho = \lambda \nu = dy = \sqrt{} ( \mathrm{R} \mathrm{R} + r r + \rho \rho )  \\
& \pi \sigma = \lambda o = dz = \sqrt{} ( \mathrm{S} \mathrm{S} + s s + \sigma \sigma ) \\
& \Phi \rho = \mu \nu = \sqrt{} (dx ^2 + dy^2 ) = \\
&
\sqrt( (\mathrm{Q} - \mathrm{R} )^2 + 
(q - r)^2 + (\Phi - \rho )^2 )   \\
& \Phi \sigma = \mu o = \sqrt{} (dx ^2 + dz^2 ) = \\
&
\sqrt( (\mathrm{Q} - \mathrm{S} )^2 + 
(q - s)^2 + (\Phi - \sigma )^2 )   \\
& \rho \sigma = \nu o = \sqrt{} (dy ^2 + dz^2 ) = \\
&
\sqrt( (\mathrm{R} - \mathrm{S} )^2 + 
(r - s)^2 + (\rho - \sigma )^2 ) ,  
\end{split}
\end{equation*} 
where we  applied the values taken from $\S.~32$. 

{\bf 76}.~~~Then the three latter equations, combined with the former, are 
reduced to these:
\begin{equation*}
\mathrm{Q} \mathrm{R} + q r + \Phi \rho = 0 ; \  \mathrm{Q} \mathrm{S} + q s + \Phi \sigma = 0 
\ \mathrm{and} \  \mathrm{R} \mathrm{S} + r s + \rho \sigma = 0 .
\end{equation*}
Moreover, if the values assigned in $\S.~34$ are substituted for the letters
$\mathrm{Q}$, $\mathrm{R} $, $\mathrm{S} $, $q$, $r$, $s$, $\Phi $, $\rho $,
$\sigma $ and the vanishing terms for the rests are neglected, the three
former will give 
\begin{equation*}
\begin{split}
& 1 = 1 + 2 \mathrm{L} dt ; \quad l + \mathrm{M} = 0   ; \\
& 1 = 1 + 2 m dt ; \quad \lambda + \mathrm{N} = 0      ; \\
& 1 = 1 + 2 \nu dt ; \quad \mu + n = 0 ,
\end{split}
\end{equation*}
so that we have $\mathrm{L} = 0 $ $m = 0$ and $\nu = 0 $, $\mathrm{M} = - l $, $\mathrm{N} = - \lambda $
and $n = - \mu $. 

{\bf 77}.~~~Thus, the three velocities of this point $\lambda $ would have to be
compared to the condition that the following hold\footnote{In the printed
  version, but not in Euler, 1752,  there are several sign mistakes.}
\begin{equation*}
\begin{split}
& du =  l d y + \lambda dz ; \\
& dv = -l d x + \mu dz     ;\\ 
& dw = - \lambda d x - \mu dy  . 
\end{split}
\end{equation*}
But the second condition demands a motion of the fluid such that 
$l = \mathrm{M} $, $\lambda = \mathrm{N} $ and $n = \mu $; hence all the
coefficients $l$, $\lambda $ and $\mu $ vanish; also the
velocities $u$, $v$ and $w$ will take the same value everywhere in the fluid. 
Therefore it is plain that the motion of the fluid
cannot correspond to solid-body motion other than pure translational.

{\bf 78}.~~~To ascertain the effect of the forces which act from the outside
    upon the fluid, it is first necessary to
    determine those forces\footnote{Here, internal forces are meant.} 
    which are required for effecting the motion
    which we have assumed to exist in the fluid. These are
    equivalent to the forces which in fact work upon the fluid;
    furthermore we have seen above in $\S.~56$ that three
    accelerating forces are required, which are here repeated. If an
    element of fluid is conceived here, whose volume, or mass is $dx
    dy dz$, the moving forces required for the motion are
\begin{equation*}
\begin{split}
& \mathrm{par.} \ \mathrm{A} \mathrm{L} = 2 dx dy dz ( \mathrm{L} u + l v + \lambda w + \mathfrak{L} ) = \\
& 
2 dx dy dz \left( u \frac{du}{dx} + v \frac{du}{dy} + w \frac{du}{dz} + 
\frac{du}{dt} \right) ; \\ 
& \mathrm{par.} \ \mathrm{A} \mathrm{B} = 2 dx dy dz ( \mathrm{M} u + m v + \mu w + \mathfrak{M} ) = \\
&
2 dx dy dz \left( u \frac{dv}{dx} + v \frac{dv}{dy} + w \frac{dv}{dz} + \frac{dv}{dt} \right) ; \\ 
& \mathrm{par.} \ \mathrm{A} \mathrm{C} = 2 dx dy dz ( \mathrm{N} u + n v + \nu w + \mathfrak{N} ) = \\
&
2 dx dy dz \left( u \frac{dw}{dx} + v \frac{dw}{dy} + w \frac{dw}{dz} + \frac{dw}{dt} \right) , 
\end{split}
\end{equation*}
so that by triple integration the components of the total forces which
must act on the whole mass of fluid may be obtained.

{\bf 79}.~~~But since the second condition requires that $u dx + v dy + w dz $
be a complete differential, whose integral is $\mathrm{S}$, let us put as before,
with time allowed to vary, $d \mathrm{S} = u d x + v dy + w dz +
\mathrm{U} dt 
$. Since $\frac{du}{dy} = \frac{dv}{dx} $; $\frac{du}{dz} =
\frac{dw}{dx} $; $\frac{du}{dt} = \frac{d\mathrm{U} }{dx} $ those
three moving forces emerge:\footnote{There is a misprint in the printed
  version, $w$ instead of $+$.}
\begin{equation*}
\begin{split}
& \mathrm{par.} \ \mathrm{A} \mathrm{L} = 2 dx dy dz \left( \frac{ u du + v dv + w dw + d\mathrm{U} }{dx} 
\right)  \\
& \mathrm{par.} \ \mathrm{A} \mathrm{B} = 2 dx dy dz \left( \frac{ u du + v dv + w dw + d\mathrm{U} }{dy} 
\right)  \\
& \mathrm{par.} \ \mathrm{A} \mathrm{L} = 2 dx dy dz \left( \frac{ u du + v dv + w dw + d\mathrm{U} }{dz} 
\right)  
\end{split}
\end{equation*}

{\bf 80}.~~~Let us set now $u u + v v + w w + 2 \mathrm{U} = \mathrm{T} $.  The
function $\mathrm{T}$ depends on  the coordinates $x$. $y$, $z$; take it at a given 
instant of time $t$:\footnote{There is a misprint: $u$ instead of $\kappa$.} 
\begin{equation*}
d \mathrm{T} = \mathrm{K} dx + k dy + \kappa dz .
\end{equation*}
The three moving forces of the element $dx dy dz $ are\footnote{Here is again a misprint: $k$ instead of $\kappa $.}
\begin{equation*}
\begin{split}
& \mathrm{par.} \ \mathrm{A} \mathrm{L} = \mathrm{K}  dx dy dz  \\
& \mathrm{par.} \ \mathrm{A} \mathrm{B} = k dx dy dz \\ 
& \mathrm{par.} \  \mathrm{A} \mathrm{C} = \kappa dx dy dz 
\end{split}
\end{equation*}
and by triple integration these formulas ought to be extended throughout
the mass of the fluid; thus forces equivalent to all\footnote{The
  pressure forces.} and their directions may be obtained. Truly
this discussion is for a later investigation, which I shall not deepen
here.

{\bf 81}.~~~Furthermore, the quantity $\mathrm{T} = u u + v v + ww + 2
\mathrm{U} $, which is analyzed in this calculation, furnishes a simpler
formula for expressing the pressure through the height $p$; we have
indeed $p = \mathrm{C} - z - \mathrm{T} $ when the particles
of the fluid are pressed upon solely by the gravity.  But if an
arbitrary particle $\lambda $ is acted upon by three accelerating
forces which are $\mathrm{Q} $, $\mathrm{q} $ and $\Phi $, acting
parallel to the directions of the axes $\mathrm{A} \mathrm{F} $, $
\mathrm{A} \mathrm{B} $ and $\mathrm{A} \mathrm{C} $, respectively,
after a calculation similar to the previous one has been carried out, the pressure
will be given by
\begin{equation*}
p = \mathrm{C} + \int ( \mathrm{Q} dx + q dy + \Phi dz ) - \mathrm{T} .
\end{equation*}  
Thus it is plain that the differential $ \mathrm{Q} + q dy + \Phi dz $
must be complete, as otherwise a state of equilibrium, or at least a
possible one, could not exist. That this condition must be imposed on
the acting forces $\mathrm{Q} $, $q$ and $\Phi $ was shown very
clearly by the most famous Mr. Clairaut.\footnote{Clairaut, 1743.}

{\bf 82}.~~~Here are, therefore, the principles of the entire doctrine of the
    motion of fluids, which, even if they at first sight may seem
    insufficiently fruitful, nevertheless embrace almost everything
    treated both in hydrostatics and in hydraulics, so that these
    principles must be regarded as having very broad extent. For this
    to appear more clearly, it is worthwhile to show how
    the precepts learned in hydrostatics and hydraulics follow.

{\bf 83}.~~~Let us therefore consider first a fluid in a state of rest, so that
    we have $u = 0$, $v = 0$ and $w = 0$;  in view of $\mathrm{T} =
    2 \mathrm{U} $, the pressure in an
    arbitrary point $\lambda $ of the fluid is
\begin{equation*}
p = \mathrm{C} + \int (\mathrm{Q} dx + q dy + \Phi dz ) - 2 \mathrm{U}.
\end{equation*}
Here, $\mathrm{U} $ is a function of the time $t $ itself which we take
as constant. Indeed, we investigate the pressure at a given time; the
quantity $\mathrm{U} $ can be included in the constant $\mathrm{C} $, so
that we obtain
\begin{equation*}
p = \mathrm{C} + \int (\mathrm{Q} dx + q dy + \Phi dz )
\end{equation*}
where $\mathrm{Q} $, $q$ an $\Phi $ are the forces acting on the particle of
water $\lambda $, parallel to the axes $\mathrm{A} \mathrm{L} $, $\mathrm{A}
\mathrm{B} $ and $\mathrm{A} \mathrm{C} $.

{\bf 84}.~~~The pressure $p$ can only depend on the position of the
point $\lambda $ that is on the coordinates $x$, $y$ and $z$; it is
thus necessary that $\int (\mathrm{Q} dx + q dy + \Phi dz )$ be a
prescribed function of them, which therefore admits
integration. Thus it is firstly clear that in the manner indicated
the fluid cannot be sustained in equilibrium, unless the forces acting
on each element of the fluid are such that the differential
formula $\mathrm{Q} dx + q dy + \Phi dz $ is complete.
Thus, if its integral is denoted $\mathrm{P} $, the
pressure at $\lambda $ will be $p = \mathrm{C} + \mathrm{P} $. Therefore,
if the only force present is gravity, impelling parallel to the
direction $\mathrm{C} \mathrm{A} $, we shall have $p = \mathrm{C} - z
$; hence, if the pressure is fixed at one point $\lambda $,
the constant $\mathrm{C} $ can be obtained. From which  the pressure at
a given time will be defined completely at all points of the fluid.

{\bf 85}.~~~However, with time passing, the pressure at a given place can
change;  and this plainly occurs, if variability is assumed 
for the forces impelling
on the water, whose calculation cannot be made from  those forces which are
assumed to act on each element of the fluid,\footnote{That is the internal
  pressure forces.}
but in such a way that they keep each other in equilibrium and
produce no motion. But if, moreover, these forces are not subjected to
any change, the letter $\mathrm{C} $ will indeed denote a constant
quantity, not depending on time $t$; and at a given location $\lambda
$ we will always find the same pressure $p = \mathrm{C} + \mathrm{P}
$.

{\bf 86}.~~~It is possible to determine the the extremal shape of a fluid in a
permanent state, when it is subjected to no forces.\footnote{Here, Euler will
comment on the shape of the free (extreme) surface of a fluid contained in an
open vessel.} Certainly, at the extreme surface of the fluid at which the
fluid is left to itself and not contained within the walls of the vase in
which it is enclosed, the pressure must be zero. Thus we shall obtain the
following equation: $\mathrm{P} = \mathrm{const} $; the shape of the external
surface of the fluid is then expressed through a relation between the three
coordinates $x$, $y$ and $z$. And if for the external circumference held
$\mathrm{P} = \mathrm{E} $, since  $\mathrm{C} = - \mathrm{E} $, in
another arbitrary internal location $\lambda $ the pressure would be $p =
\mathrm{P} - \mathrm{E} $.  In this manner, if the particles of the fluid are
driven by gravity only, and because $p = \mathrm{C} - z $, the following will
hold at for the external surface $z = \mathrm{C} $; from which the external
free surface is perceived to be horizontal.

{\bf 87}.~~~Next, everything which has so far been brought out concerning
     the motion of a fluid through tubes is easily derived from these
     principles. The tubes are usually regarded as very narrow, or
     else are assumed to be such that through any section normal to
     the tube the fluid flows across with equal motion: from there
     originates the rule, that the speed of the fluid at any place in
     the tube is reciprocally proportional to its amplitude. Let
     therefore $\lambda $ be an arbitrary point of such a tube, of
     which the figure is expressed by two equations relating the three
     coordinates $x$, $y$ and $z$, so that thereupon for any abscissa
     $x$ the two remaining coordinates $y$ and $z$ can be defined.

{\bf 88}.~~~Let henceforth the cross section of this tube at $\lambda$ be  
$r r $; in
another fixed location of the tube, where the cross secion is $ f f $,
let the velocity at the present time be $ \taurus $; now after the 
time $dt$ has elapsed, let the velocity  become $ \taurus + d \taurus $, so that
$\taurus $ is a function of time $t$, and similarly with
$\frac{d\taurus }{dt} $. Hence the true velocity of the fluid at
$\lambda $ will be at the present time $V = \frac{ff \taurus }{rr} 
$. Since now $y$ and $z$ are obtained from the shape of the tube, we have
$dy = \eta dx $ and $dz = \theta dx $; thus the three
velocities of the point $\lambda $ in the fluid, parallel to
directions $\mathrm{A} \mathrm{L} $, $\mathrm{A} \mathrm{B} $ and
$\mathrm{A} \mathrm{C} $, are
\begin{equation*}
\begin{split}
& u = \frac{ff \taurus }{r r} \, \frac{1}{\sqrt{(} 1 + \eta \eta + \theta \theta )} ; \  
v = \frac{ff \taurus }{r r} \, \frac{\eta }{\sqrt{(} 1 + \eta \eta + \theta \theta )} ; \\
& w = \frac{ff \taurus }{r r} \, \frac{\theta }{\sqrt{(} 1 + \eta \eta + \theta \theta )} , \  
\end{split}
\end{equation*}
and hence, $ u u + v v + w w = \mathrm{V} \mathrm{V} = \frac{
f^4 \taurus \taurus }{r^4 } $: and $r r $ is function of $x$ itself,
thus of the dependent variables $y$ and $z$.

{\bf 89}.~~~Since $u dx + v dy + w dz $ must be a complete differential, the integral
    of which is denoted  $=S$, we have:
\begin{equation*}
\begin{split}
d \mathrm{S} = \frac{ff \taurus }{r r} \, \frac{dx (1 + \eta \eta +
\theta \theta ) }{ \sqrt{(} 1 + \eta \eta + \theta \theta ) } =
\frac{ff \taurus }{r r} \, dx \sqrt{(} 1 + \eta \eta + \theta \theta )
.
\end{split}
\end{equation*}
Moreover, $dx \sqrt{(} 1 + \eta \eta + \theta \theta ) $ expresses the
element of the tube itself; if we denote it by $ ds$, we shall obtain
$d\mathrm{S} = \frac{ff \taurus ds}{rr} $: although $\taurus $ is a function of
the time,\footnote{As was stated in \S.~88.} here we fix the time and, 
furthermore,
the quantities $s$ and $rr$ do not depend
on time but only on the shape of the tube; thus we have
$\mathrm{S} = \taurus \int
\frac{ff ds }{rr} $.

{\bf 90}.~~~Turning now to the pressure $p$ which is found at the point
of the tube $\lambda $, the quantity $\mathrm{U} $ has to be
considered;  it arises from the differentiation of the quantity
$\mathrm{S} $, if the time only is considered as variable, so that we
have $ \mathrm{U} = \frac{d\mathrm{S}}{dt} $.  Thus, since the integral
formula $\int \frac{ff ds}{rr} $ does not involve time $t$, on the
one hand we shall have $\frac{d\mathrm{S} }{dt} = \mathrm{U} =
\frac{d\mathrm{U} }{dt } \int \frac{ff ds }{r r} $, and on the other
hand it will follow from \S.~80 that:
 \begin{equation*}
 \mathrm{T} = \frac{f^4 \taurus \taurus }{r^4 } + \frac{2 d\taurus }{dt } 
\int \frac{ff ds}{r r} .
 \end{equation*}
Therefore, after introducing arbitrary actions of forces $Q$, $q$ and $\Phi $, 
the pressure at $\lambda $ will be
\begin{equation*}
\begin{split}
p = \mathrm{C} + \int (Q \, dx + q\, dy + \Phi \, dz ) - \frac{f^4 \taurus \taurus }{r^4 } - 
\frac{2 d \taurus }{dt } \int \frac{ff ds }{rr}  
\end{split}
\end{equation*}
This is that same formula which is commonly written for the
motion of a fluid through tubes; but now much more widely valid, since
arbitrary forces acting on the fluid are assumed here, while commonly
this formula is restricted to gravity alone. Meanwhile it is in order
to remember that the three forces $\mathrm{Q}$, $q$ and $\Phi$ must be such
that the differential 
formula $Q \, dx + q \, dy + \Phi \, dz $ be complete,
that is, admit integration.

\end{document}